\documentclass{IEEEtran} 
\ifCLASSINFOpdf
\else
\fi
\usepackage{multirow}
\usepackage{array}
\usepackage{booktabs}
\usepackage{graphicx}
\usepackage{xcolor}
\begin{document}
%
% paper title
% Titles are generally capitalized except for words such as a, an, and, as,
% at, but, by, for, in, nor, of, on, or, the, to and up, which are usually
% not capitalized unless they are the first or last word of the title.
% Linebreaks \\ can be used within to get better formatting as desired.
% Do not put math or special symbols in the title.
\title{SmartSeed: Smart Seed Generation for Efficient Fuzzing}
%
%
% author names and IEEE memberships
% note positions of commas and nonbreaking spaces ( ~ ) LaTeX will not break
% a structure at a ~ so this keeps an author's name from being broken across
% two lines.
% use \thanks{} to gain access to the first footnote area
% a separate \thanks must be used for each paragraph as LaTeX2e's \thanks
% was not built to handle multiple paragraphs
%

\author{{Chenyang~Lyu,~Shouling~Ji,~Yuwei~Li,~Junfeng~Zhou,~Jianhai~Chen,~Jing~Chen}% <-this % stops a space
\thanks{C. Lyu, S. Ji, Y. Li, J. Zhou and J. Chen are with the College of Computer Science and Technology, Zhejiang University, China. E-mail: \{puppet, sji, liyuwei, zhoujf620, chenjh919\}@zju.edu.cn.}
%\thanks{P. Zhou is with the School of Electronic  Information and Communications, Huazhong University of Science and Technology,  Wuhan, Hubei, 430074, China. E-mail: panzhou@hust.edu.cn.}
\thanks{Jing Chen is with the College of Computer Science and Technology, Wuhan University, China. E-mail: chenjing@whu.edu.cn.}
}

\maketitle

% As a general rule, do not put math, special symbols or citations
% in the abstract or keywords.
\begin{abstract}
Fuzzing is an automated application vulnerability detection method. For genetic algorithm-based fuzzing, it can mutate the seed files provided by users to obtain a number of inputs, which are then used to test the objective application in order to trigger potential crashes. 
As shown in existing literature,   
the seed file selection is crucial for the efficiency of fuzzing. 
However, current seed selection strategies do not seem   to be   better than randomly picking seed files. 
Therefore, in this paper, we propose a novel and generic system, named \emph{SmartSeed}, to generate seed files towards efficient fuzzing.  Specifically, \texttt{SmartSeed} is designed based on a machine learning model to learn and generate high-value binary seeds.  We evaluate \texttt{SmartSeed} along with American Fuzzy Lop (AFL)   on 12  open-source applications with the input formats of mp3, bmp or flv. 
We also combine \texttt{SmartSeed} with different fuzzing tools to examine its  compatibility.   From extensive experiments, we find that \texttt{SmartSeed} has the following advantages: 
First, it only requires tens of seconds  to generate sufficient high-value  seeds. 
 Second,  it   can generate seeds with multiple kinds of input formats and significantly improves the   fuzzing  performance  for most applications with the same input format.     
Third, \texttt{SmartSeed} is compatible to  different fuzzing tools.  
In total, our system discovers more than twice unique crashes and   5,040 extra unique paths than the existing best seed selection strategy  for the evaluated 12 applications. 
From the crashes found by \texttt{SmartSeed}, we discover   16 new vulnerabilities and have received their CVE IDs. 
\end{abstract}

% Note that keywords are not normally used for peerreview papers.
\begin{IEEEkeywords}
Fuzzing, vulnerability detection, seed generation.
\end{IEEEkeywords}

% For peer review papers, you can put extra information on the cover
% page as needed:
% \ifCLASSOPTIONpeerreview
% \begin{center} \bfseries EDICS Category: 3-BBND \end{center}
% \fi
%
% For peerreview papers, this IEEEtran command inserts a page break and
% creates the second title. It will be ignored for other modes.
\IEEEpeerreviewmaketitle

\section{Introduction}\label{section1}

An application bug refers to an error, failure or fault in a computer system's application (such as an operating system or an image browser) that causes it to behave wrong. There are several common application bug types such as buffer overflow, integer overflow,  use-after-free and so on.

Application bugs can cause enormous harms. 
An application bug of Therac-25 resulted in the deaths of several patients \cite{item12}. 
%An air crash happened and all the passengers died because of the bug in engine softwares \cite{item13}. 
%Intel paid \$475 million for the bug of  Pentium processors in 1,995 \cite{item14}. 
On June 4, 1996, an Ariane 5 rocket exploded just forty seconds after its lift-off. It was caused by an application error and lost the cargo which is worth \$500 million \cite{item20}. 
In 2002, a study by the National Institute of Standards and Technology (NIST) estimated that application bugs costed U.S. \$59.5 billion loss a year \cite{item26}. 
Thus, application bugs can cause both security problems and financial loss. It is important to find and fix application bugs before they cause accidents or are exploited by  attackers. 

Over the decades, extensive research has been spent for application bug detection. 
%There are several popular methods to detect undiscovered bugs because of the pressing needs, 
There are several existing staple  techniques in the academia and industry community, 
such as   dynamic taint analysis, symbolic execution and fuzzing.  
%Generally speaking, existing detection methods can be categorized into three classes: static analysis, dynamic taint analysis, and fuzzing: 
%there are several popular methods to detect undiscovered bugs on account of the pressing needs: 

%Static analysis is a way of detecting bugs without actually executing softwares \cite{item15, item16}. It detects vulnerabilities by scanning the code of the program. A static analysis tool can analyze the control flow or the data flow of a software. Thus, static analysis tools have high execution efficiency. However, the tools have a high false positive rate. Researchers still have to spend a lot of time to check the code. 

Dynamic taint analysis is one of the most popular methods to detect application bugs \cite{item17, item18}.  It marks the data from untrusted inputs as tainted.  Then, the analysis tool traces the tracks of all the tainted data when the application is running. If the tainted data is used in some dangerous memory,  the analysis considers that a bug is detected. Dynamic taint analysis has a low false positive rate by discovering  bugs in this way.  However,  since the analysis needs to add necessary checks and traces the track of  data when the application is running,  dynamic taint analysis   has low execution efficiency.

Symbolic execution is  a method of detecting bugs by considering the whole process as solving constraints \cite{item27, item28}.  It regards the input of an objective application as variable X and tries to figure  out the relationship between X and the execution paths of the application. If a   symbolic execution tool runs into a branch statement, it will record two  constraints, one for each branch. Thus, when symbolic execution is finished, users have a bunch of constraints, each of which corresponds to an execution path of the objective application. If users want to analyze one of the paths, they can solve  the corresponding constraint to get the corresponding X.  In other words, users obtain the corresponding input to test the target execution path. Symbolic execution can accurately locate the bugs in an application. However, it has a poor scalability. When  the objective application is large, the final equation becomes too  complicated to be solved for.

Fuzzing  is one of the most effective methods to detect crashes in applications. Different from other kinds of   methods, fuzzing dose not analyze which  data or which code results in the crashes. A fuzzing tool uses a simple yet efficient way to detect crashes. It generates a great number of input files. Then, the tool tests the objective application with these inputs and detects whether the application behaves abnormally. If the objective application happens to have a crash, the fuzzing tool will store the input file that triggers the crash. In this way, the fuzzing tool can detect the crashes and obtain the input files which trigger  crashes. Then,  users can study these input files to figure out whether the objective application has  bugs. 
However, since  fuzzing  does not consider how to trigger  crashes in an objective application, the whole process of detecting crashes is blind. %It is hard to say that the strategy of the fuzzing tool is high-efficiency. 
Fortunately, with the help of current powerful computing capability, a fuzzing tool can use  a large number of input files to test an  objective application so many times within a short time. Therefore, fuzzing can potentially find more crashes. %while other methods are analyzing the control-flow or the codes of the softwares. 
In this paper, we focus on providing a better seed set for   fuzzing   in order to discover more crashes. 

%Nowadays, since  softwares are getting bigger and more complex, the efficiency of fuzzing tools is getting harder and harder to meet users' needs. 
%Generally, to improve  fuzzing efficiency, we basically have two directions: design better fuzzing tools and use a better fuzzing seed set. 

Nowadays, modern application is much larger and more complex, which makes it harder to adapt the fuzzing tools as needed. To improve fuzzing efficiency, researchers typically develop strategies for improvement following two directions: design better fuzzing tools and use a better fuzzing seed set.

Following the first direction, there are a number of works focusing on designing better tools or improving current fuzzing tools. 
%The fuzzing tools can be classified into two categories: (1) the fuzzing tools based on generation rules; (2) the fuzzing tools based on genetic algorithms. 

The fuzzing tools based on generation rules can learn the input format of the objective application \cite{ item6, item19, item24, item25,  item31}. Then, they can generate highly-structured input files based on the input format.  
While other fuzzing tools spend most time passing the format check,  generation-based fuzzing tools use the generated highly-structured files to test the execution of   applications. 
Thus, the generation-based fuzzing tools are better at detecting the crashes of applications which check the syntax features and the semantic rules of the input files. 
%The generation-based fuzzing tools use the generative highly-structured input files to test the objective software and watch whether the software has a crash. 

The fuzzing tools based on genetic algorithms do not consider the input format of the objective application.   According to  genetic algorithms, the tools mutate the initial input seed set by byte flipping, cross over and so on. Then, in order to discover   crashes, the tools take the initial input seed set and the mutated input files as the inputs of the applications. 
%If an input file detects a new path or a new crash of the objective software, the mutation-based fuzzing tool will store the input file locally. What's more, the mutation-based fuzzing tool will add this input file into the seed set. Then, the child input files will be generated from this input file according to the genetic algorithm. 
Since they require little prior knowledge of the objective application, the mutation-based fuzzing tools work pretty efficiently. However, sometimes they get stuck because of the simple crash detection strategy. 
To improve the efficiency of mutation-based fuzzing tools, a number of researches combine the fuzzing tools with other vulnerability detection technologies such as static analysis, taint analysis and symbolic execution \cite{item4, item5, item7,  item22, item23}.  
There are several other  researches indicating that stimulating fuzzing tools to improve the coverage and test low-frequency paths can improve the efficiency of fuzzing \cite{item2, item3, item45, item46}, where new fuzzing tools based on their viewpoints are presented respectively.

%Haller et al. presented a new fuzzing tool named \emph{Dowser}, which combines taint tracking, program analysis and symbolic execution to focus on the buffer overflow and underflow bugs \cite{item7}. 
%Sang et al. combined fuzzing with white-box symbolic analysis \cite{item4}. Then, the new tool found 38.6\% more bugs on average than the previous fuzzing tools during the same time. 
%Stephens et al. combined fuzzing with selective concolic execution and presented the new fuzzing tool named \emph{Driller} \cite{item5}. When a path seems interesting but fuzzing cannot generate the input files to enter the path, \emph{Driller} will use selective concolic execution to explore this path. 
%There are several researches indicated that stimulating fuzzing tools to test low-frequency paths can improve the efficiency of fuzzing \cite{item2, item3}. 
%Rebert et al. pointed out that a better seed selection can improve the efficiency of fuzzing \cite{item1}. 
%Wang et al. presented a probabilistic model named PCSG to generate highly-structured input files \cite{item6}. It improves the effectiveness of fuzzing softwares such as XML engines. 
%However, to our knowledge, there is no research that using a machine learning model to generate valuable input files for fuzzing tools. 

The second direction of improving the efficiency of  fuzzing tools is to use a better seed set. 
Allen and Foote presented an algorithm to consider the parameter selection and automated selection of seed files \cite{item30}. The algorithm was implemented in an open-source Basic Fuzzing  Framework (BFF). 
Woo et al. developed an analytic framework to evaluate 26 randomized online scheduling algorithms that schedule a better seed to fuzz a program \cite{item29}. 
%One of their new scheduling algorithms performed the best during their experiments. 
In 2014, Rebert et al. evaluated six seed selection strategies of fuzzing and showed how to select the best selection strategy \cite{item1}, etc.

%In 2,014, Rebert et al. presented several interesting and impressive conclusions of how to select the seed set \cite{item1}. First, a better input file seed set can improve the performance of fuzzing tools. Second, a better input file seed set can be  transferable. It can improve the performance of fuzzing softwares with the same input format. Third, it is hard to say that the current selection algorithms are strictly better than others. Sometimes they perform very close to randomly selecting the seed set. Fourth, computing a better seed set for a given software costs a lot. It may be unworthy to compute a seed set for a single software. 
%The benefits may not exceed the cost. 
%However, it may be worthy to compute a better seed set for one input format. The seed set will be used when fuzzing the softwares with the same input format. 

However, the current seed selection strategies have many deficiencies. 
For example, some  strategies require a lot of time to obtain the seed set \cite{item1}. 
What's more, proved by our experiments and the abovementioned work \cite{item1}, the current seed selection strategies perform unstable in many application scenarios. Further,  they do not have evident advantages than random seed selection in many cases. 

%Sometimes the selected seed set performs worse than randomly selecting the files as the seed set. 
%One of the problems is that researchers can obtain a great many of input files from the Internet. Then, they use the algorithms to select a better input seed set from the huge set, which requires a lot of time. 
%Another problem is that the current problems seem not to perform strictly better than selecting the seed set randomly. 

%They cost a lot but seem not to perform better than selecting the seed randomly. One of the problems is that researchers can obtain a great many of input files from the Internet. Then, they use the algorithms to select a better input seed set from the huge set, which requires a lot of computing power. 

%Since it costs a lot to select a better seed set from the regular files and the algorithms perform badly, 

Therefore, to solve the problem of how to obtain a better seed set for the applications without highly-structured input format, we come up with the following heuristic questions: 

%\begin{enumerate}
%\item \textbf{Can we use machine learning algorithms to effectively generate an input seed set?} 
%It is inapposite to spend a lot of time on obtaining the seed set for fuzzing. Thus, one of our targets is to generate the set fast. 
%Is generating a seed set faster than the current seed selection strategies?  

%Can we use machine learning algorithms to generate a better seed set, rather than select from the huge input set? Can the generated seed set improve the efficiency of fuzzing tools? 

%\item \textbf{Does the generated input seed set have robustness?} 
%It is inefficient if we have to run the whole system to obtain the seed set  while fuzzing the new software. Therefore, one of our objectives is that  the generated input seed set can improve the most softwares with the same input format once we finish the training of our system. 
%Does the generated set have the best performance on the most of the softwares with the same input format? 
\textbf{Q1: Can we generate valuable seeds in a fast and effective manner?}  
As we discussed before, many existing seed selection strategies are slow, which is improper for fuzzing.   
More importantly, existing solutions cannot yield effective seeds in many scenarios.   
Therefore, to address these limitations, instead of studying how to select seeds, our primary goal is to study how     to automatically generate effective seeds in a fast manner leveraging   state-of-the-art machine learning techniques.    

\textbf{Q2: Can we generate valuable seeds in a robust manner?}   
Based on the assumption that we have already figured out a fast and effective   seed generation strategy, it is still inefficient if \texttt{SmartSeed} can only generate valuable seeds for some specific input format or  
we have to retrain the     model everytime we want to fuzz a new application   with the same input format.  
Thus, our second goal is to design a robust seed generation system.    
It should be able to generate valuable seeds for multiple input formats.   
Moreover, we only need to train the model once for any kind of input format. Then, the files generated by this model can improve the fuzzing performance  for other applications with this input format.    %What's more, SmartSeed can generate valuable seeds for multiple input formats. 
% the generated seed files of which will improve the performance  while fuzzing the most softwares with the same input format. 

\textbf{Q3: Can we generate valuable seeds in a compatible manner?}  
It is unexpected for most fuzzing cases if our system  can  only  generate valuable seeds for specific fuzzing  tools.   
Therefore, to improve the compatibility, we aim to design a   seed generation  strategy that   can combine with different fuzzing tools and improve their performance.

%\item \textbf{Does this system have  generality?} 
%We aim to design a system that not only can  generate the seed sets with different input formats, but also can combine with different fuzzing tools. 
%Can the system deal with different input formats? Or can the system combine with different fuzzing tools? 
%\end{enumerate}

\begin{figure}[h] 
\begin{center} 
\includegraphics[width=0.35\textwidth,height=1.5in]{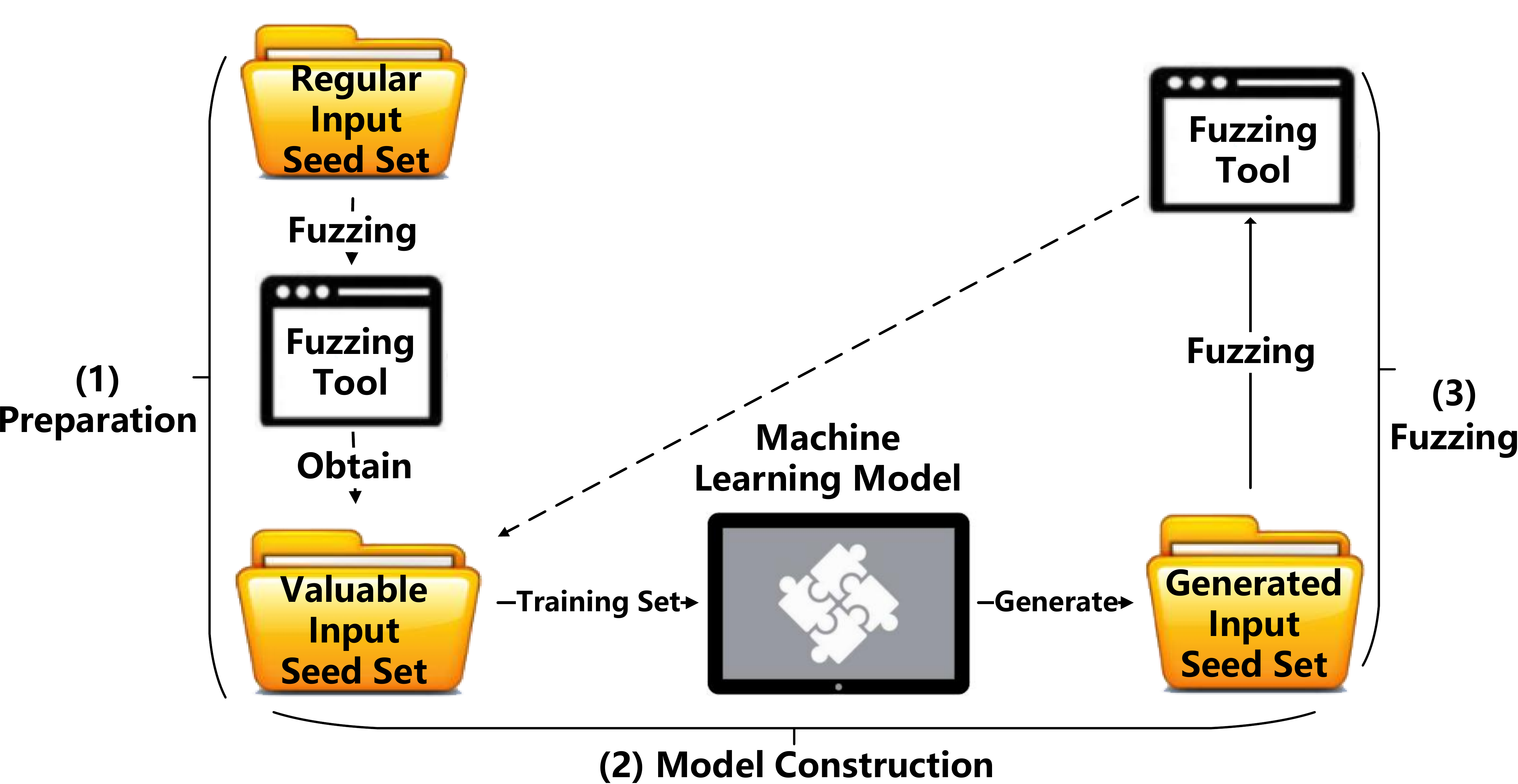}
\caption{Workflow of   \texttt{SmartSeed}.}\label{fig1}
\end{center} 
\end{figure}

Following the above heuristic questions,   
  in this paper, we present a novel seed generation system named \emph{SmartSeed} to provide  fuzzing tools with a better seed set  as shown in Fig. \ref{fig1}.

Basically, the workflow of  \texttt{SmartSeed} consists of three stages. 
%\begin{enumerate}
%\item In order to construct an initial valuable  input seed set to bootstrap SmartSeed generation, we use regular files to fuzz the  softwares and collect the input files triggering the unique crashes and paths. 
%\item We use the valuable input seed set as the training set to train the machine learning model of SmartSeed. Then, use the model to generate  similar files as the training set. 
%\item Fuzzing tools use the generated  files as the input seed set to fuzz an objective software. 

(1)  \textbf{Preparation.} \texttt{SmartSeed} is a machine learning-based system. To bootstrap   \texttt{SmartSeed}, we need to prepare necessary training data. Specifically, we collect some regular files and employ them to fuzz some applications by commonly used fuzzing tools like American Fuzzy Lop (AFL) \cite{item11}. Then, we collect the input files that trigger unique crashes or new paths as the training data. Note that, this step is only for collecting necessary training data to bootstrap 
  \texttt{SmartSeed} (which can then be used to generate seeds for fuzzing many applications), and can be easily implemented in practice. We show the details in Section \ref{trainset}.  

(2)   \textbf{Model Construction.} To make \texttt{SmartSeed} easily extendable in practice, we propose a transformation mechanism to encode the raw training data into generic matrices, which are then being  employed to construct a generative model for seed generation.  
Leveraging the generative model, we   generate effective files as seeds.  

(3)   \textbf{Fuzzing.}  Leveraging  the seeds generated from the constructed generative model, we  use fuzzing tools (e.g., AFL) to discover    crashes of objective applications.   
%\end{enumerate}

Note that the whole process can form a closed-loop. 
The machine learning model can generate  effective seed files to help  fuzzing tools discover  new crashes and paths of objective applications. Then, the training set of the machine learning model in   \texttt{SmartSeed} can be refined and enhanced with complementary input files that trigger new crashes or paths. 
%Thus, if the whole process is the mutual promotion process, the users can excavate more and deeper crashes of the objective software by doing the whole process repeatedly. 

With the help of \texttt{SmartSeed}, users can efficiently generate a valuable input seed set.   
Our evaluation on 12 open source applications demonstrates that \texttt{SmartSeed} significantly improves the performance of fuzzing   compared to state-of-the-art seed selection strategies.  
%In most cases, the seed set from our system can significantly improve the performance of fuzzing tools as shown in our experiments.
%In this paper, %we introduce a novel way to acquire a number of valuable input flies for efficient fuzzing. 
The main contributions  in this paper can be summarized as follows: 

$\bullet$
We present a machine learning-based system named \texttt{SmartSeed} to generate valuable binary seed files for fuzzing  applications without of  requiring highly-structured input format.  
% or input format check. 

$\bullet$
Combining with AFL, we evaluate the seed files generated by \texttt{SmartSeed} on 12 open source applications with the input formats such as mp3, bmp or flv.  Compared with   state-of-the-art seed selection strategies,   \texttt{SmartSeed} finds  608 extra unique crashes and 5,040 extra new paths than the existing best strategy in total.   

$\bullet$
We further combine \texttt{SmartSeed} with other popular fuzzing tools to examine its compatibility:   
(1) combining with AFLFast \cite{item2}, \texttt{SmartSeed} is still the best seed strategy for each application.   
(2) \texttt{SmartSeed + honggfuzz} \cite{item34} finds the most crashes for five of the  six objective applications.  (3) Among the  evaluated seed selection/generation strategies,   
only \texttt{SmartSeed} discovers   crashes on \texttt{ps2ts} and \texttt{mp42aac}  when using VUzzer \cite{item3} as the fuzzing tool.   
%We evaluate the generative model by using the generated input files to fuzz the softwares with the same input format. The results show that the generated input files have the following advantages: first, these input files can improve the performance of fuzzing tools on the softwares which provide the training set of the generative model. Second, the input files have transportability. They can improve the performance on a large proportion of softwares with the same input format. 

%$\bullet$ 

$\bullet$
%We analyze the reasons why the performance of the generated input files is better. In the end, 
We further analyze the seed sets generated by \texttt{SmartSeed} and  other state-of-the-art seed selection strategies,  
and present several interesting findings to enlighten the research of fuzzing.  
Visualized by t-SNE \cite{item35},  
the seed files generated by \texttt{SmartSeed} are closer to the most valuable files that trigger crashes or paths.   Meanwhile, the files of \texttt{SmartSeed} that trigger   unique crashes cover the largest area, which  implies   that the generated files are easier to be mutated into more discrete valuable files.   
What's more, we realize that the execution speed is an improper indicator for discovering crashes. However, the   larger generation of seeds   helps fuzzing tools  discover more unique paths.   
%(CVE). 
In total, \texttt{SmartSeed}  finds 23 unique vulnerabilities on 9 applications, including 16 undiscovered ones.  
In the end, we open source the code of \texttt{SmartSeed}, which is expected to  facilitate  future fuzzing research \footnote{We also make our system publicly available to facilitate the research in this area, which is available at: https://github.com/puppet-meteor/SmartSeed.}. 

%\textcolor{red}{
The remainder of this paper is organized as follows. 
%In Section \ref{section2}, we introduce the background of our work. 
In Section \ref{section3}, we describe the detail of \texttt{SmartSeed}. 
We evaluate \texttt{SmartSeed} and   existing state-of-the-art seed selection strategies on  12 applications, combine \texttt{SmartSeed} with different fuzzing tools and analyze the results of vulnerabilities discovered by different seed selection strategies  in Section \ref{section4}.   
In Section \ref{section9}, we further analyze the performance of different seed selection strategies. 
In Section \ref{section5}, we make some discussions and remark the limitation of our work.  %In Section \ref{section6}, we overview the related work. 
We conclude the paper in Section \ref{section7}.  
We provide more evaluation  results and summarize the related work with  remarks    in the \emph{Supplementary File}.

\section{SmartSeed}\label{section3}

%In this section, we elaborate the system of \texttt{SmartSeed}. 

\subsection{System Architecture}

The core idea of \texttt{SmartSeed} is   to construct a generative model. Then, we use this   model to fast    generate valuable files as the input seed set of  fuzzing tools.   

\begin{figure}[h] 
\begin{center} 

\includegraphics[width=0.35\textwidth,height=1.6in]{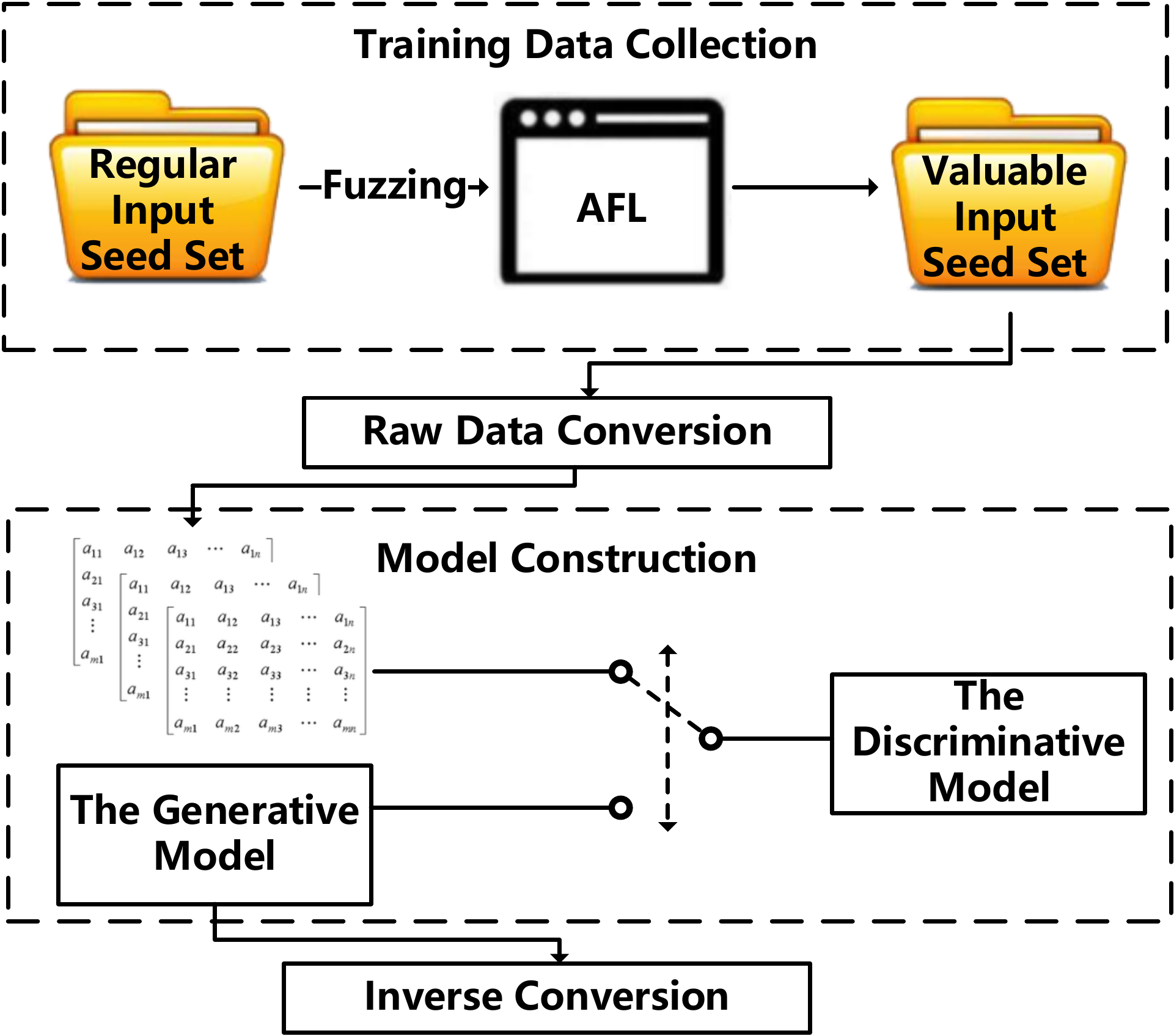}

\caption{Architecture of \texttt{SmartSeed}.}\label{fig2}

\end{center} 
\end{figure}

%As shown in Fig. \ref{fig2}, 
%Overall, the whole process of constructing the generative model and generating    seeds   is shown as Fig. \ref{fig2}. 
As shown in Fig. \ref{fig2}, the whole architecture of \texttt{SmartSeed}  can be divided into 4 procedures:  
%\begin{enumerate}

(1) \textbf{Training data collection:} We introduce a   criterion to measure the value of input files   and present a method to obtain a  training set for \texttt{SmartSeed} (Section \ref{trainset}).  
%Obtain the training set, which will be elaborated in Subsection \ref{trainset}. 

(2) \textbf{Raw data conversion:} To deal with files with unfixed formats or unfixed file sizes, we convert the binary files of   raw training data to a uniform  type of matrices (Section \ref{Conversion}).  \label{step2} 

(3)  \textbf{Model  construction:}  
Taking the matrices as training data, we  construct a seed  generative model based on Wasserstein Generative    Adversarial Networks   (Section \ref{model}).   
%Use the matrices to be the training set of  Wasserstein GAN (WGAN) model and train the model \cite{item10}. 

(4) \textbf{Inverse conversion:} Based on the generative model, we generate new matrices and    convert them into proper input files, which is the reverse process of   Procedure (2) (Section \ref{inverse}). 

 %To be specific, restore the elements of the matrix to the large numbers. Convert the number of decimal system to six numbers of Base65 system. Convert the number of Base65 system to the character of Base64. Decode the character string of Base64 into the binary file and store it locally. 
%\end{enumerate}

%It is worth to note that users of SmartSeed can select an alternative machine learning model as the generative model according to  their application scenarios. In our implementation, we use WGAN   to generate the seed set. The reason will be introduced in Subsection \ref{model}. 

In our system, the employed fuzzing tool   can    be flexible, i.e.,    
\texttt{SmartSeed} can be combined with most existing mutation-based fuzzing tools.    
Since   AFL  is one of the most efficient existing fuzzing tools \cite{item11},   
%many researches admit AFL and regard it as the benchmark fuzzing tool.    
%Thus, 
 by default, we select AFL to be the fuzzing tool in our implementation.    
%American Fuzzy Lop (AFL) is one of the most popular existing fuzzing tools \cite{item11}. The main procedures of AFL are as follows: 

%\begin{enumerate}
%\item Users of AFL prepare input files that satisfy the input format of the objective software. 

%\item AFL regards these input files as seed files. According to the genetic algorithm, AFL mutates each file of the seed files by byte flipping and so on. Then, AFL obtains a great sum of input files. 

%\item AFL uses these input files to run the objective software ceaselessly and watches the behavior of the software. If AFL finds a crash or a new path of the objective software, it will save the corresponding input file locally. In addition, the corresponding input file will be regarded as a new seed file. Therefore, more input files generated based on this new seed file. 
%\end{enumerate}

%In the following subsections, we mainly elaborate how to obtain the training set and how to convert the training set to the appropriate expression form for WGAN model. 

\subsection{Training Data Collection} \label{trainset}  
%In this subsection, we explain how to obtain the training set of WGAN model. 

To construct a machine learning model for generating valuable seed files,   
 we need to   obtain an initial training set  first.  
 %the first step is   to obtain an initial training set for constructing the model. 
%As we decide to use the machine learning model to generate valuable seed files, the first problem   is how to obtain the training set of the model. 

%We deem that the input files in the training set should be valuable, i.e., these files are supposed to be more effective for  discovering the unique crashes and paths. Otherwise,  SmartSeed may not be able to learn the features of   valuable files that can improve the fuzzing performance. 
%In other words, the input file should detect a new path or a new bug of the softwares. 
%Then how to evaluate whether an input file is valuable? 

%Since the ultimate goal of  fuzzing  is to detect more crashes, the input files are considered as valuable if they can trigger  crashes of  objective softwares. What's more, existing research proves that increasing the coverage and the depth of   fuzzing paths are more likely to  increase the number of   explored crashes \cite{item2, item3}. Thus, we consider an input file is   valuable if it can detect a new path of the objective software. 
%As a conclusion, we define ``valuable input files'' as the input files that trigger  unique crashes or  unique paths of  softwares. 

Certainly, we are expected to ensure that an input file in the training set is really valuable.   
%An input file in the training set is valuable means the file is supposed to be more effective for discovering the unique crashes and paths. 
Otherwise,  \texttt{SmartSeed} may not be able to learn useful features of  ``valuable input files'' and further generate such kind of files.   
Therefore, we first clarify valuable input files.   
Specifically, in our implementation, we define  valuable files as the input files that trigger unique crashes or    unique paths of applications. The reasons are as follows:   
%In order to collect valuable input files to train an effective generative model, 
 (1) since the ultimate goal of  fuzzing  is to detect more crashes, the input files are considered as valuable if they can trigger unique crashes of objective applications; (2) according to the existing research   \cite{item2, item3, item45, item46}, increasing  the coverage and the depth of fuzzing paths are more likely to increase the number of explored crashes. 
 Hence, the files triggering new paths are also valuable from this perspective. 
%We define ``valuable input files'' as the input files that trigger unique crashes or unique paths of softwares. 

Intuitively, we may employ   existing seed selection strategies to select a few valuable input files as the training set.   
However, according to the  existing research \cite{item1} and the results of our experiments,   current seed   selection strategies seem to be unreliable.  
%Thus,  using   current seed selection strategies is difficult to select  valuable seed files from the huge input files, which are usually downloaded from the Internet. 
Then, we realize that fuzzing tools such as AFL will store the   input files that trigger unique crashes or paths,   which  faultlessly coincides with our needs.  
Thus, we have the following training data collection strategy:   
we can first use   regular input files collected from the Internet to fuzz the applications with the same input format.  
Then,   
we gather the valuable input files, which trigger   unique paths or   unique crashes of those  applications, as the   training set of \texttt{SmartSeed}.  % that trigger the unique paths or the unique crashes of the softwares. 

\begin{figure*}[h] 
\begin{center} 
\includegraphics[width=0.94\linewidth,height=1in]{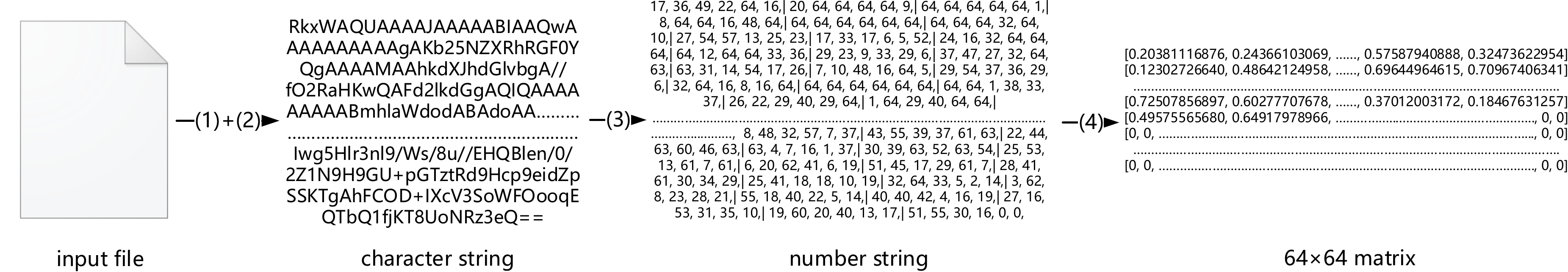}
\caption{Training Data Conversion.}\label{fig4}
\end{center} 
\end{figure*}

%Since we can run as many AFL fuzzing procedures  at the same time as the CPU resources we have, 
Note that, the above training set construction is not a limitation of    \texttt{SmartSeed} in practice. First, the   criterion for determining a valuable input file is that it can trigger a crash or a new path, which can be easily   implemented by AFL in practice.   
Second, for the applications with the same input formats, such as the applications that deal with mp3, we can fuzz   multiple applications simultaneously to facilitate the training set construction process, and meanwhile increase   
  the multiformity of the input files in the training set.   
Third, since we can parallelly run  many   fuzzing procedures,   
we can further accelerate the training set collection process, e.g., we can collect more than 20,000 valuable files   for one kind of input format within a week.   
%we collect more than 20,000 files as the training set   for one input format within one week. 
Therefore, we can construct the training set for \texttt{SmartSeed} easily and fast in practice.   

By collecting a training set in this way, we further have the following advantages: 

$\bullet$ 
We can accurately evaluate the value of the input files. The input files in the training set can certainly detect  unique paths or trigger   unique crashes of some applications. 
Thus, they carry useful features for learning. 

$\bullet$ 
During the fuzzing process, we realize that the formats of  many files that trigger   crashes or new paths are  corrupted. We analyze the reason is that files are randomly mutated according to the employed genetic  algorithm, and it seems that the corrupted files are more likely to trigger unique crashes and paths. However, it is    difficult to gather a number of corrupted files from the Internet, while \texttt{SmartSeed} can be trained to   generate many corrupted files as expected.   
%Generated by the genetic algorithm, most valuable input files are mutated randomly and corrupted. It is difficult to gather a number of corrupted files from the Internet. Whereas, we can easily collect the corrupted files by fuzzing the softwares. 

\subsection{Raw Data Conversion}\label{Conversion}

%Now we consider how to convert the training files to matrices for the machine learning model. 

%Since most files in the training set are corrupted and one of our objectives is to make SmartSeed deal with multiple input formats, 
%we should figure out a conversion method that can convert the training files with unfixed  file sizes and unfixed formats. 

%WGAN model cannot read the different files of the training set in respective ways. 

%Based on the knowledge that  machine learning algorithms are better to work on   quantitative values of   matrices rather than some specific value types, we are expected to find a way that can convert multiple types of files to a uniform type of matrices. 
%In addition, we hope to give expression to the magic bytes in the binary form of   training files. Then, a machine learning model would be easier to learn the features of   magic bytes. Thus, the files should be read as binary form and be converted to the matrices. 

To construct a generic seed generation model, we propose a mechanism to convert the raw input files in the   training set to a uniform type of matrices. The reasons for conducting such conversion are as follows.  

First, one of our objectives is to make \texttt{SmartSeed} deal with multiple input formats and unfixed  file   sizes. However, it is inconvenient to adjust the data read mode for different kinds of and different sizes of files.   Thus, we should figure out a uniform method to read data from the training set.  
Second, the formats of  many files in the training set are corrupted. A normal reading manner, such as reading a  bmp picture file as a  three-dimensional matrix, may not work in many application scenarios.  
Third, based on the knowledge that  machine learning algorithms are better to work on   quantitative values of   matrices rather than some random value types, we are expected to find a way that can convert multiple types of files to a uniform type of matrices.  
Finally, we expect to give expression to the magic bytes in the binary form of   training files. Because in this way   a machine learning model would be easier to learn the features of   magic bytes that control the code execution   paths. Thus, the files should be read in binary form and are expected to be converted to   uniform matrices.

Below, we introduce the main procedures    of raw training data conversion, which are shown in Fig. \ref{fig4}.   
%As a result, the main procedures we convert files to matrices are as follows: 
%\begin{enumerate}

(1)  Since all the files can be read in binary form,  we can read the binary form of  any type of   file and get a binary string. 

(2)  To   handle the problem of how to recognize the end of   the binary string, we encode the  string with Base64. Thus,   we have the string formed by 64 kinds of   characters and ``='', e.g., the   \emph{character string}   shown  in Fig. \ref{fig4}. 

\begin{figure}[h] 
\begin{center} 

\includegraphics[width=0.35\textwidth,height=1.2in]{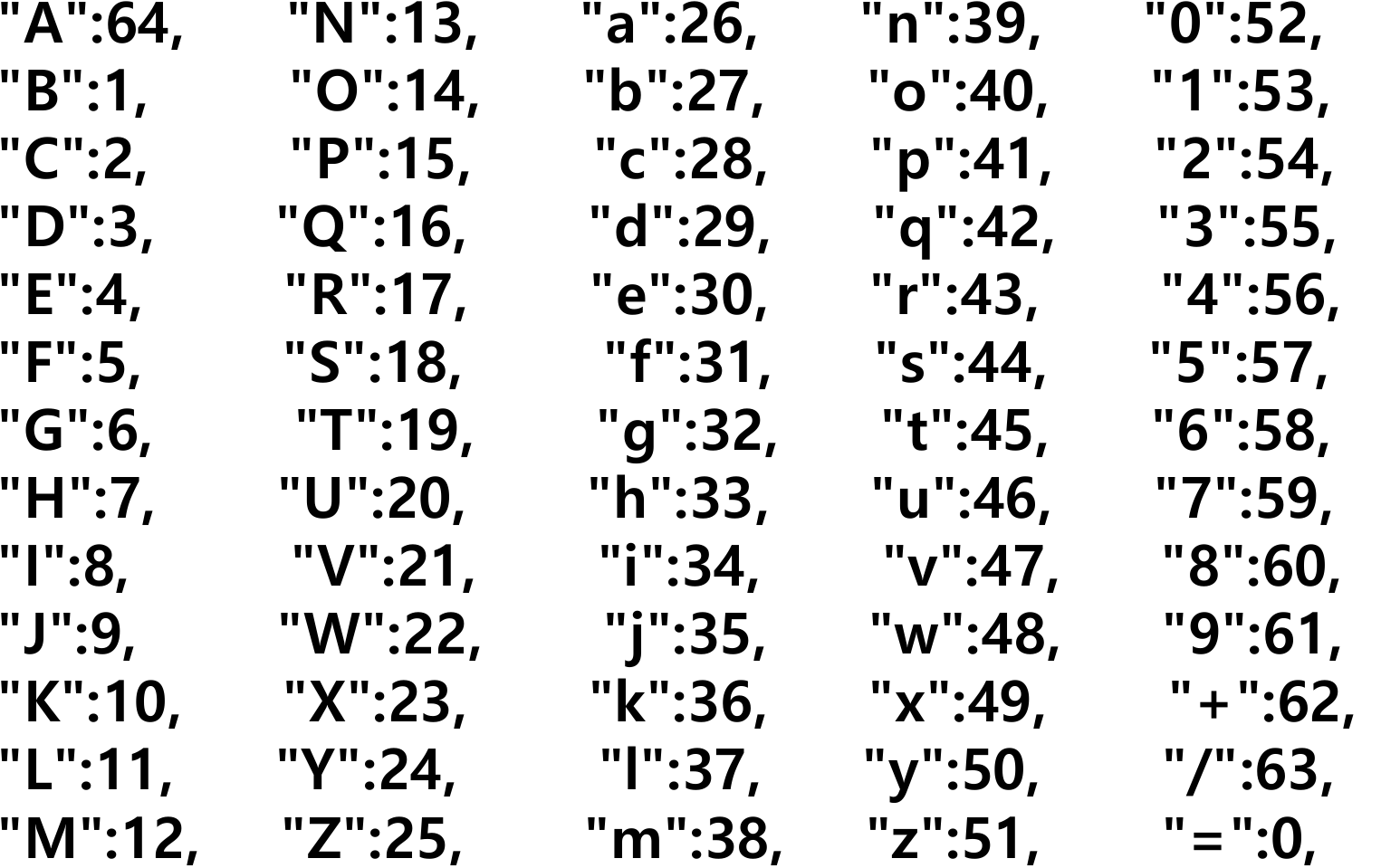}

\caption{Convert Base64  Character  to the Number from 0 to 64.}\label{fig3}

\end{center} 
\end{figure}

(3)  Now,  since a  character string  may have 65 different characters, for convenience, we convert the characters of Base64 and ``=''  to the numbers  (from 0 to 64). Thus, we obtain a string of numbers as   shown in Fig. \ref{fig4}. The    encoding mechanism for such conversion is shown in Fig. \ref{fig3}.  

(4)  To economize   the number of   elements in a matrix, we convert every six numbers of the number string to   a large number of the decimal system. Then, we normalize the numbers of  the decimal system to [0,  0.75418890624]  (since $65^6-1=75,418,890,624$)   for the    accuracy   and efficiency of the model training. Finally, add 0 at the end of the matrix if there is an empty element, as shown in  Fig. \ref{fig4}.  
%\end{enumerate}

Since the larger numbers will be stored in the scientific notation form and lose some veracity, one element of the matrix can only store one number in the decimal system that is converted from six numbers  (from 0 to 64)  at most. However, users can choose the number less than six if the file size is small in practical application   scenarios.    
Note that the fewer numbers are converted into one large number, the easier a machine learning model learns the features of input files. We convert 6 numbers in our experiments in order to test the most difficult  situation for the machine learning model. 

%By convert raw input files in this method, we have the following advantages:
Leveraging the above method,  we can convert raw files with corrupted formats or unfixed file sizes into a   uniform type of matrices.   
%$\bullet$
This can significantly improve the  extendability and  compatibility of \texttt{SmartSeed}.

%During our experiment, it seems that how to convert the character strings to the matrices has less impact on the training process. We analyze the reason is as follows: WGAN model works on the quantitative value of the matrices. No matter how you convert the character string, the influence is that the quantitative value of the matrix changes. WGAN model seems just learn the features that change from one quantitative value to another quantitative value. 

\subsection{Model Construction} \label{model}

%It is worth to note that users of SmartSeed can select an alternative machine learning model as the generative model according to  their application scenarios. 

One of the best existing generative models is the Generative Adversarial Networks (GAN) model,  
which has been widely used for unsupervised learning since 2014  \cite{item8}.  
GAN   is a new framework consisting of a generative model and a discriminative model.  
The generative model tries to generate   fake data that are   similar to the real data in the training set, while the   discriminative model tries to distinguish the fake data from the real data. Two models alternatively work   together to train  each other and further to improve each other. As a result, the generative model  will generate   data that are too real to be distinguished by  the discriminative model. 
Usually, the generative model provided by GAN can generate more realistic data than other algorithms.   However, it is unstable to train a GAN model. GAN also has many problems such as model collapse.   
%In other words, GAN model only generates one or two types of files learned from the training set. 

Radford et al. tried multiple combinations of   machine learning schemes to construct better  generative and   discriminative models for  GAN. They presented the Deep Convolutional GAN (DCGAN) model   \cite{item9}. DCGAN   can generate more realistic data than standard GAN  and  is much easier for training.   However, it still has many problems including model collapse.  

In 2017, Arjovsky et al. presented the Wasserstein GAN (WGAN) model \cite{item10}. Unlike other GAN models,  WGAN   improves the stability of learning a lot. It is also much easier to train a WGAN model. In   addition, WGAN   can solve the problems of GAN like mode collapse in most application scenarios. 

Hence, for our purpose, we employ WGAN to learn the characteristics of valuable   files and then generate   valuable   seed files.  
%In other words, if it needs to satisfy several magic bytes to enter the paths, the magic bytes will be embodied in the binary mode of the files. Then, WGAN model may learn this feature and can generate the files satisfying the magic bytes. 
%In addition, since mutation-based fuzzing tools always damage the file format, it is acceptable that WGAN model cannot generate the files with the rigorous file format. 
%Thus, we decide to use WGAN   to learn from the training set and generate our seed set.  
%In the end, we briefly introduce the training process of WGAN model. 
Our selection further has the following benefits.   
First,  WGAN   can learn the features of the training set by itself.   
Thus, we do not need to pay attention to the feature selection, which is very time-efficient.   
Another advantage is that we can freely choose an appropriate machine learning model as the generative model  and the discriminative model of WGAN.  
We analyze that Multi-Layer Perceptron (MLP)  focuses more on every quantitative value in a matrix, while  Convolutional Neural Network (CNN) pays more attention to the global features of a  matrix.  
Then, in order to construct a better model, we test the performance of both  neural network models. 
For our application, MLP does work better than CNN as the generative and   discriminative models of WGAN.   
It also requires less training time. 
Therefore, we choose MLP as the model in \texttt{SmartSeed}. 
Now, based on the collected training data, we can train a generative model for valuable seed generation.

Note that,  a detailed description of  WGAN can be found in  \cite{item10}.   Since our focus in this paper is to construct a GAN-based efficient fuzzing framework and further demonstrates its effectiveness, we leave the research of developing an improved GAN model as the future work.   
Furthermore, users of \texttt{SmartSeed} can select an alternative machine learning model as the generative model in terms of the application scenarios.

\subsection{Inverse Conversion}\label{inverse}
In this subsection, we   introduce how to employ  the generative model of   \texttt{SmartSeed}   to obtain an   effective  input seed set.   
Since the training set of \texttt{SmartSeed}   is a number of matrices, the generative model is trained to  generate similar matrices. To obtain binary  input files   for fuzzing, we have to convert the generated matrices to   binary files.  
Thus, we do the reverse of the abovementioned procedures in Section \ref{Conversion}.  

To be specific, the fist step is to restore the [0,  0.75418890624] elements of the matrix to   large numbers of the decimal system.  
Second, convert each number of the decimal system to six numbers (from 0 to 64), i.e., the  \emph{number string} as shown in Fig. \ref{fig4}.  
Then, convert the numbers (from 0 to 64) to the characters of Base64 and ``='', i.e., the    \emph{character string} as shown in Fig.  \ref{fig4}.  
Finally, decode the character string of Base64 into a binary file and store it locally.  
Thereout, we obtain the   input files  for the following fuzzing. 

%Since the size of the files in the training set is between 12KB and 16KB, to control the irrelevant variable, we consider that SmartSeed should generate the files with the similar file size. Therefore, we implement SmartSeed to generate the generative file larger than 11KB. Users can adjust the parameter to generate the files with an appropriate file size. 

\section{Implementation \& Evaluation}\label{section4}

%In this section, we first implement SmartSeed and five state-of-the-art seed selection strategies as the input set of AFL in order to watch whether our system improves the performance of AFL. 
%Of course, users can use any other kind of the mutation-based fuzzing tools. We choose AFL since it is one of the benchmark fuzzing tools and is admissive by the researchers. 

%Then, we combine SmartSeed and other two baseline seed selection strategies with different fuzzing tools in order to see whether our system has the generality. 

%In this section, we implement \texttt{SmartSeed} and compare its performance with   state-of-the-art seed selection strategies. 

\subsection{Datasets}

To evaluate the performance of \texttt{SmartSeed}, we employ 12 open-source linux applications with input formats of   mp3, bmp or flv as shown in Table \ref{tabletarget}.

\begin{table}[!hbp]

\centering
\caption{Objective applications.}\label{tabletarget}
%\begin{tabular}{|c|c|c|c|}
\begin{tabular}{ p{0.4cm}<{\centering} p{0.9cm}<{\centering} p{0.7cm}<{\centering}  p{1.8cm}<{\centering} p{2.9cm}<{\centering}} %{| p{0.9cm} <{\centering}| p{0.9cm} <{\centering}| p{1.2cm} <{\centering}| p{1.2cm} <{\centering}|}
%\toprule
\toprule[1pt] 
  & Target  & Version & Release Time & Heat \\
\midrule[1pt]
\multirow{4}{0.8cm}{mp3} 
    & mp3gain &1.5.2.r2&2010-08&639 downloads/week\\
    & ffmpeg &3.4&2018-02&10.6k collections\\
    & mpg123 &1.25.6&2017-08&1,720 downloads/week\\
    & mpg321 &0.3.2&2012-03&78 downloads/week\\
\hline
\multirow{4}{0.8cm}{bmp} 
    & magick   &7.0.7&2017-10&1.7k collections\\
    & bmp2tiff   &3.8.2&2007-03&52 collections\\
    & exiv2 &0.26&2017-05&67 collections\\
    & sam2p &0.49.4&2017-12&10 collections\\
    \hline
    \multirow{4}{0.8cm}{flv} 
    & avconv   &12.2&2017-09&576 collections\\
    & flvmeta &1.2.1&2016-08&67 collections\\
    & ps2ts   &1.13&2015-11&70 collections\\
    & mp42aac  &1.5.1.0&2017-09&443 collections\\
%\hline
\bottomrule[1pt]  
%12.5\% & 1,536 & 8/32 & 12.44\% & 38.65\% & 0.36\% \\
%\hline
%\bottomrule
\end{tabular}
\end{table}

%\texttt{mp3gain}  is a popular  tool to analyze and adjust the volume of mp3 files. From \emph{SOURCE FORGE}, we learn that all versions of \texttt{mp3gain} for linux have been downloaded   639 times in total this week. Note that   all versions of \texttt{mp3gain} for Windows has been downloaded  10,639 times in total this week.  

%\texttt{ffmpeg}   is a collection of libraries and tools to process multimedia content such as audio, video, subtitles and related metadata, which has been collected 10.6k times in \emph{github}. 

%\texttt{magick}  is the main program of \texttt{ImageMagick}, whose mirror has been collected 1.7k times in \emph{github}. %Some research also uses \texttt{ImageMagick} as the objective software \cite{item36}. 

%\texttt{bmp2tiff}  is the conversion tool of \texttt{libtiff}, whose mirror has been collected 52 times in \emph{github}. Some research also uses \texttt{libtiff} as the objective software \cite{item36}. 

%As for flv, \texttt{avconv} and \texttt{ps2ts} are the media tools  provided \texttt{libav} and \texttt{tstools}, respectively. Although \texttt{mp42aac}  is the tool in \texttt{Bento4} that deal with mp4 files,  we use flv files as the initial seed set and discover the crashes. 

We choose these applications mainly for three reasons.  

First, they can be fuzzed with the files without high-structured formats, which are the focus of genetic algorithm-based fuzzing as well as \texttt{SmartSeed}.  

Second, these applications are popular and important   open-source   programs. For instance, \texttt{mp3gain} is a   popular tool to analyze and adjust the volume of mp3 files. From \emph{SourceForge} \cite{item37},   \texttt{mp3gain} for linux are downloaded 639 times on average per week and  \texttt{mp3gain} for Windows  are downloaded 10,877 times on average per week.    %\texttt{ffmpeg} is a collection of libraries and tools to  process multimedia content such as audio, video, subtitles and related metadata, which has been collected 10.6k  times on \emph{github}.   
\texttt{magick} is the main program of \texttt{ImageMagick}, whose mirror has  been collected 1.7k times on \emph{github}. \texttt{bmp2tiff} is the conversion tool of \texttt{libtiff}, which is employed in the experiments of many  other researches as a fuzzing dataset \cite{item36}. As for flv,  \texttt{avconv} and \texttt{ps2ts} are the popular media tools provided by \texttt{libav} and \texttt{tstools},   respectively.  

%\textcolor{red}{
Third, these applications have different  code logic and functionalities, and thus are sufficiently representative. For example, \texttt{magick} is used to browse bmp, \texttt{sam2p} is used to convert bmp into eps, \texttt{bmp2tiff} is used to convert bmp into tiff, and \texttt{exiv2} is a cross-platform C++ library and a command line utility to manage image metadata. All the four applications are provided by different groups.  
Although \texttt{mp42aac} is a tool in \texttt{Bento4} that deals with mp4 files, we also use  \texttt{SmartSeed} to generate flv files as the initial seed set to fuzz it. This is mainly for evaluating the  robustness of \texttt{SmartSeed}.

\subsection{SmartSeed Implementation}

To implement \texttt{SmartSeed}, the first step is to collect the training set. 
As we mentioned before, we   fuzz some applications and collect the valuable input files.

In consideration of the training efficiency, 
if the size of training files is too large, we have to use a big  matrix to store the file, which leads to  longer  training time. On the contrary, it is hard to collect small files for meaningful multimedia data with formats such as mp3 and flv. What's more, small  files may not carry enough features for machine learning.  
Therefore, we need to determine a proper size for the employed training files.  

Considering both the training efficiency and the training effectiveness, we prefer to employ files less than 17KB to construct the generative model of \texttt{SmartSeed}. To accommodate files with size of 17KB or less, $64  \times 64$ matrices are sufficient (whose maximum storage is 18KB).  
On the other hand,  it would be hard for \texttt{SmartSeed} to learn meaningful features from too sparse   matrices, e.g., a matrix with more than 35\% of its elements are null, since they carry less valuable information. 
 Therefore, we finally decide to collect valuable input files with size between 12KB and 17KB as the training data in our implementation.  
This setting can ensure that  the training rate is fast, make it easy to collect valuable files  and meanwhile ensure  sufficient information is carried, i.e., seek a balance between   efficiency and utility.  

Note that   easy    extendability  is one of \texttt{SmartSeed}'s advantages.  The  size  of the input  files in the training set is  adjustable in practice. If users want to use   smaller files as the training set, they may  convert fewer characters of Base64 to the number  of the decimal system or use   smaller matrices.  
On the other hand, if users want to use files with bigger size to train \texttt{SmartSeed}, they may employ  a bigger matrix  to store the string of Base64.

Now, we are ready to collect the training data. The collection processes are conducted on 16 same virtual  machines with an Intel i7 CPU,  4.5GB memory   and a Ubuntu 16.04 LTS system and   last for a week  for  each input format. For applications with the mp3 format, we employ AFL to fuzz \texttt{mp3gain} and  \texttt{ffmpeg}, and collect 20,646 valuable input files with size 12KB to 17KB as the training set.  
For bmp, we use AFL to fuzz \texttt{magick} and \texttt{bmp2tiff}. Then, we collect 24,101 input files as the  training set.  
For applications with the flv format,  we employ AFL to fuzz \texttt{avconv} and \texttt{flvmeta}. Then, we  collect 21,688 input files between 12KB and 17KB that discover   new paths or   unique crashes.

Based on the collected training data, we construct the prototype of \texttt{SmartSeed}. The core function  of \texttt{SmartSeed} is  implemented in Python 2.7. The code of the WGAN model is implemented in  Pytorch 0.3.0. 
We use Adam as the optimization algorithm of  the  WGAN model.    %which is one of the best   optimization algorithms. 
Then, we test several times to decide the hyper-parameters, such as how many times to train the discriminative model after training the generative model. During the training process, we decrease the learning rate of the  optimization algorithm from $0.5 \times 10^{-3}$  to $0.5 \times 10^{-12}$  inch by inch.  
It takes us around 38 hours to train the generative model of \texttt{SmartSeed} for 100,000 times on a server   with 2 Intel Xeon E5-2640 v4 CPUs running at 2.40GHz,  64 GB memory, 4TB HDD and   a GeForce GTX 1080TI GPU card. Note that others may not need to train the models for 100,000 times. % in our experiments we choose to train 100,000 times. 

\subsection{Effectiveness} \label{six}  % and Analysis}

\subsubsection{Comparison Strategies} 
%\subsubsubsection{Seed   Strategies Presentation}
To compare the performance of \texttt{SmartSeed} with   state-of-the-art seed selection strategies, we  implement the following methods. 

%\textbf{SmartSeed.} \textbf{SmartSeed} is the seed set that generated by our system. The system learns from the training set and generates the 100 similar fake files to constitute the seed set. 

\texttt{random.} Under this scheme, we randomly select   files from the regular input files that are usually  downloaded from the Internet. Because this is the most common seed selection strategy in practice, we take  \texttt{random} as the baseline seed strategy in our experiments.

\texttt{AFL-result.} Under this scheme, we randomly select seeds from the saved input files of AFL, which are  also the files used for training \texttt{SmartSeed}. Since these files  are   certainly to   either trigger   unique  crashes or new paths during the fuzzing process, they may yield an outstanding     performance as a seed set for  fuzzing tools. As our system is learnt from these files,   \texttt{AFL-result} can serve as the    baseline of  \texttt{SmartSeed}.

\texttt{peachset.} The fuzzing framework Peach provides a seed selection tool named \texttt{MinSet} \cite{item33}, whose workflow is as follows: (1)  \texttt{MinSet} inspects the coverage of each file in the full set which might be collected from the Internet; (2) it sorts  the files by their coverage    in the descending  order; (3)  
\texttt{MinSet} initializes the seed set as an empty set,  which is denoted by \texttt{peachset}; 
(4) \texttt{MinSet} checks the coverage of files in order. If a file improves the coverage of \texttt{peachset}, the file will be added to \texttt{peachset}.

\texttt{hotset.} \texttt{hotset} contains the files that discover the most unique crashes and paths within \emph{t} time \cite{item1}. To construct \texttt{hotset}, first,   use each file, that may be collected from the Internet, as the initial seed set to fuzz an application for \emph{t} seconds. Second,   sort the files in the descending  order by the number of discovered crashes and paths. Third,   select the top-\emph{k} files to constitute \texttt{hotset}. In our experiments, we fuzz each file for 240 seconds.

\texttt{AFL-cmin.} In order to select a better seed set, AFL provides a tool named \texttt{AFL-cmin}.  
The  core idea of \texttt{AFL-cmin} is to  filter out the  redundant files that inspect   already discovered paths.  
\texttt{AFL-cmin} tries to find the smallest subset of   the full set, which still has the same coverage as the full set.

\subsubsection{Results} \label{selfsame}
%\subsubsubsection{Seed   Strategies Presentation}
Now, we evaluate the effectiveness of different seed generation/selection strategies. For the schemes that need an initial dataset to bootstrap the seed selection process, we collect a dataset consisting of 4,600 input files for each input format from the Internet. For \texttt{SmartSeed}, we generate seeds based on its generative model.  All the following experiments are conducted on same machines with  an Intel i7 CPU,  4.5GB memory   and a Ubuntu 16.04 LTS system.

\textbf{Seed Generation Speed.}  
First, we evaluate the seed generation speed of different strategies. The results are shown in  Table \ref{tabletime}, from which we have the following observations.

(1)  As we discussed before, \texttt{random} simply picks a seed set at random from the initial dataset. \texttt{AFL-result} follows a similar strategy except for randomly picking a seed set from the saved input files of AFL. Therefore, they are the fastest ones for selecting seeds and are scalable.

(2)  \texttt{SmartSeed} employs a generative model to generate seeds which is also very fast and scalable. It only takes 12 seconds to generate 100 seed files and 240 seconds to generate 2,000 seed files, exhibiting a linear increasing correlationship between the time consumption and the number of generated seeds.  

(3)   For the coverage-based seed selection strategies \texttt{peachset} and \texttt{AFL-cmin}, they are relatively slow.  For instance, to generate 500 seed files, it takes \texttt{peachset} and \texttt{AFL-cmin}  2,500 seconds and 874 seconds,   respectively. This is mainly because  they have to check the coverage of each file.  Furthermore, 
since coverage-based strategies are hard to increase the number of the selected files by increasing the size of the initial file set, we have to use the screening tools to filtrate multiple initial file sets containing different files to obtain enough selected files.

(4) For \texttt{hotset},  it is extremely slow.   For instance,  \texttt{hotset}  requires  $>$2,000 min to select seeds from 500 files in our experiments.   
  This is mainly because \texttt{hotset} has to use each single file as the seed set   to fuzz an application for  sufficient time,   which is very time consuming.  
  
\begin{table}[b] 
\centering
\caption{Seed generation speed.}\label{tabletime}
%\begin{tabular}{|c|c|c|c|}
\begin{tabular}{ p{2.85cm}<{\centering} p{1.35cm}<{\centering}  p{1.45cm}<{\centering} p{1.45cm}<{\centering}} %{| p{0.9cm} <{\centering}| p{0.9cm} <{\centering}| p{1.2cm} <{\centering}| p{1.2cm} <{\centering}|}
%\toprule
\toprule[1pt] 
seed   strategy & 100 seeds &  500 seeds  & 2,000 seeds  \\ 
\midrule[1pt]
\texttt{SmartSeed} &12 sec&60 sec&240 sec\\   
\texttt{random} &0.5 sec&0.5 sec&0.5 sec\\ 
\texttt{AFL-result} &0.5 sec& 0.5 sec&0.5 sec\\    
\texttt{peachset} &   500 sec & 2,500 sec & 10,000 sec \\  
\texttt{hotset} & $>$400 min & $>$2,000 min & $>$8,000 min \\   
\texttt{AFL-cmin} &  174.8 sec &  874 sec & 3,496 sec\\
\bottomrule[1pt]  
%12.5\% & 1,536 & 8/32 & 12.44\% & 38.65\% & 0.36\% \\
%\hline
%\bottomrule
\end{tabular}
\end{table} 

In practical fuzzing applications, O(100) number of seeds are usually sufficient to bootstrap fuzzing tools. For instance, 
 3 seed files are sufficient for bootstrapping VUzzer  by default \cite{item3}.  
 Therefore, based on the results in Table \ref{tabletime}, \texttt{SmartSeed} is very efficient in generating seeds.

\begin{table*}[] 
	\centering
	\caption{Unique crashes and   paths of each objective application discovered by different fuzzing strategies. }\label{tableall}
	\begin{tabular}{p{1.2cm}<{\centering} p{0.87cm}<{\centering} p{0.87cm}<{\centering} p{0.87cm}<{\centering} p{0.87cm}<{\centering} p{0.87cm}<{\centering} p{0.87cm}<{\centering} p{0.87cm}<{\centering} p{0.87cm}<{\centering} p{0.87cm}<{\centering} p{0.87cm}<{\centering} p{0.87cm}<{\centering}  p{0.87cm}<{\centering}}  
		\toprule[1pt]
		\multirow{2}{2.2cm}{ }    & \multicolumn{2}{c}{SmartSeed+AFL} &  \multicolumn{2}{c}{random+AFL}          &  \multicolumn{2}{c}{AFL-result+AFL}  &  \multicolumn{2}{c}{peachset+AFL} &  \multicolumn{2}{c}{hotset+AFL}&  \multicolumn{2}{c}{AFL-cmin+AFL}\\ \cline{2-13} 
		                                                                  & unique crashes & unique paths & unique crashes & unique paths &  unique crashes & unique paths & unique crashes & unique paths & unique crashes & unique paths & unique crashes & unique paths     \\ \midrule[1pt]
		\multicolumn{1}{l}{mp3gain}  & \textbf{153} & \textbf{1,742} &87&936 &128&876& 129&841&119&989&95&698  \\  
		\multicolumn{1}{l}{ffmpeg} & 0 & 1,592   & 0 & \textbf{1,925}& 0 & 1,671 &0&1,129&0&1,178&0&1,306   \\  
		\multicolumn{1}{l}{mpg123}   & \textbf{78} & \textbf{2,154} &0 & 1,183 &0&1,001 & 0 & 1,405&0&1,172&0&1,589   \\  
		\multicolumn{1}{l}{mpg321} & \textbf{204} & \textbf{1,060}   & 40   & 766 & 13 & 187  & 37 & 748&16&128&72&441   \\ \hline
		\multicolumn{1}{l}{magick} & \textbf{238} & \textbf{3,374}   & 0  & 697   & 0 & 1,149   & 0  & 196&0&722 &0&265   \\  
		\multicolumn{1}{l}{bmp2tiff} & \textbf{56} & \textbf{714}  & 21   & 466  & 34 & 684  & 32  & 534 &21&583&21&498    \\  
		\multicolumn{1}{l}{exiv2} &\textbf{66}&1,549 & 20   & 1,413  & 38 & \textbf{2,293}  &55   & 1,096&57&1,593 &27&1,202    \\ 
		\multicolumn{1}{l}{sam2p} & \textbf{50} & \textbf{1,322}    & 21  & 468 & 36 & 719 & 25 & 520&28&479&12&363      \\ \hline
		\multicolumn{1}{l}{avconv}    & 0 & \textbf{4,315}     & 0   & 1,873  & 0 & 4,191 & 0  & 1,994  &0&1,976&0&2,200 \\  
		\multicolumn{1}{l}{flvmeta} & 90 & 1,259   & 68 & 886 & 87 & \textbf{1,295}  & 100 &1,104&98&1,013&\textbf{104}&1,128   \\ 
		\multicolumn{1}{l}{ps2ts} & \textbf{43} & 1,692    & 4  &  1,381  & 14 & 1,740  & 26 & 1,472&7&1,419&19&\textbf{1,742}  \\  
		\multicolumn{1}{l}{mp42aac}   & \textbf{118} &  \textbf{658}   & 80   & 329  & 102 &   585   &  84  & 571&53&338&70&453    \\ \hline
		\multicolumn{1}{l}{total}  &\textbf{1,096}&\textbf{21,431}&341&12,323&452&16,391&488&11,610&399&11,590&420&11,885\\    
		\multicolumn{1}{l}{average}&\textbf{91.3}&\textbf{1,785.9}&28.4&1,026.9&37.7&1,365.9&40.7&967.5&33.25&965.8&35&990.4  \\
\bottomrule[1pt]
	\end{tabular}
\end{table*}

\textbf{Fuzzing Effectiveness.}
Now, we evaluate the effectiveness of the seeds obtained by different strategies. In our experiments, we employ each seed generation/selection strategy to obtain a seed set consisting of 100 files and then feed the seed set to AFL.   
 Note that   it is sufficient  and effective to use 100 files as the seed set to guide fuzzing tools,  which is   analyzed in detail in the  \emph{Supplementary File}  due to the space limitations.     
%whose proof is not shown  due to the space limitations. 
To control irrelevant variables,   
we operate all the fuzzing experiments   on the same virtual machines for 72 hours with an Intel i7 CPU, 4.5G memory and a Ubuntu 16.04 LTS system.   %The number of files in each seed selection strategy is 100. 

%In order to evaluate the performance of our system, we compare the results of the six seed selection strategies while fuzzing the same software. 
The primary goal  of fuzzing   is to discover crashes. Thus, we use the number of unique crashes found by each  seed strategy as the main evaluation criterion. Furthermore, as shown in many existing researches \cite{item2,  item3, item45, item46},  a higher coverage can improve the fuzzing performance. Therefore, our second   evaluation  criterion is the number of  discovered  unique paths  during the same time.   
The results are shown in Table \ref{tableall}.  
We can deduce the following conclusions from Table \ref{tableall}.

(1)   For discovering unique crashes, \texttt{SmartSeed} is very effective and performs the best in almost all the  evaluation scenarios. When fuzzing mp3 applications, \texttt{SmartSeed} discovers 24 more crashes than the  existing best solution for \texttt{mp3gain}, and discovers more than twice unique crashes than the existing best  solution for \texttt{mpg123} and \texttt{mpg321}. When fuzzing bmp applications, \texttt{SmartSeed}  also  yields the best performance. Again, when fuzzing flv applications, \texttt{SmartSeed}   discovers the most crashes  in total among the evaluated solutions. An exception is \texttt{flvmeta}, on which   \texttt{AFL-cmin}  discovers the most crashes. 
After viewing the the saved valuable files of \texttt{flvmeta},   we conjecture the reason is that     
normal flv files are more likely to find crashes of \texttt{flvmeta}, while \texttt{SmartSeed} tends to generate  corrupted  files that are more likely to find paths of  \texttt{flvmeta}, since many of its training data are  corrupted.   
%we realize that most unique crashes of \texttt{flvmeta} still keep the format of flv, while we use the corrupted flv files, which discovered unique paths, to train \texttt{SmartSeed}. This reason can also explain the performance of \texttt{AFL-result}.    

(2)    For triggering unique paths, again, \texttt{SmartSeed} is very effective. Among the 12 evaluated applications,  \texttt{SmartSeed} discovers the most paths on 8 ones, the second-most paths on one and the third-most paths  on three ones. 
Specifically, for mp3, \texttt{SmartSeed} discovers nearly   twice unique paths than other fuzzing strategies  except for \texttt{ffmpeg}.  
 It also outperforms other    strategies when fuzzing \texttt{magick}, \texttt{bmp2tiff} and \texttt{sam2p}.  %\texttt{AFL-result + AFL} discovered the most paths of \texttt{exiv2}. 
  Both \texttt{SmartSeed}  and  \texttt{AFL-result}  perform  well when fuzzing    flv   applications.

(3)  
Interestingly, for \texttt{mpg123} and \texttt{magick}, \texttt{SmartSeed + AFL} is the only one that can  discover crashes among the evaluated strategies.   Specifically, it discovers 78 unique crashes on \texttt{mpg123}  (mainly bus errors) and 238 crashes on \texttt{magick} (mainly segmentation faults). This again suggests that  \texttt{SmartSeed} is very effective in practice. Moreover, for \texttt{ffmpeg} and \texttt{avconv}, no strategy  finds any crash. We conjecture the reasons are: these two applications are very robust and/or our fuzzing time  might be too short to trigger crashes.

%(3)  
%The input seed set generated by \texttt{SmartSeed} distinctly improved the performance of fuzzing the objective softwares in most cases, it found more unique crashes and paths than other seed selection strategies. 
%In total, \texttt{SmartSeed + AFL} discovered  608 extra unique crashes than the  second fuzzing strategy, which is more than twice than the second best strategy. \texttt{SmartSeed + AFL} discovered  5,040 extra unique paths than the second seed selection strategy. 
%On average, \texttt{SmartSeed + AFL} discovered 124.6\% more unique crashes and 30.7\% more unique paths than the current best seed selection strategy in 72 hours. 

(4)  
For \texttt{random}, \texttt{AFL-result}, \texttt{peachset}, \texttt{hotset} and \texttt{AFL-cmin}, it is  difficult to say which   is better. In total, the unique crashes and paths discovered by them are around [340, 490]  and [11,000, 16,500], respectively.    
   \texttt{peachset} seems to perform better than others in more scenarios. However, even for the naive solution  \texttt{random}, it   outperforms others in many applications. For instance, \texttt{random} discovers more  crashes than \texttt{peachset} and \texttt{AFL-result} for \texttt{mpg321},  discovers more crashes for  \texttt{mpg321} and \texttt{mp42aac} and the same number of crashes for \texttt{bmp2tiff} compared with  \texttt{hotset}, and discovers more crashes for \texttt{sam2p} and \texttt{mp42aac} and the same number of  crashes for \texttt{bmp2tiff} compared with \texttt{AFL-cmin}, which is  unexpected.

%(4)  
%Only \texttt{SmartSeed + AFL} discovered  bus error on \texttt{mpg123}  and segmentation fault on \texttt{magick}.  However, no fuzzing strategy discovered the crashes on \texttt{ffmpeg} or \texttt{avconv}. 
%It is hard to say that which is the best seed selection strategy in the other five seed selection strategies. \texttt{peachset} seems outperform the other four seed selection strategies  by a very narrow margin. \texttt{random} performs the worst on the most objective softwares. 
%Although  \texttt{peachset} and \texttt{AFL-cmin} are the seed selection strategies based on the coverage of files, they only discover more unique paths than \texttt{hotset} and are worse than the baseline \texttt{random}, which is unexpected. 

%In summary, 
%\texttt{SmartSeed} outperform the other seed selection strategies on the numbers of both unique crashes and paths when combing with AFL to  fuzz softwares with  input formats of mp3, bmp and flv. 

In summary, compared with state-of-the-art seed selection strategies,    \texttt{SmartSeed} is more stable and  yields much better performance. In   total, \texttt{SmartSeed + AFL} discovers 124.6\% more unique crashes  (i.e., 608 extra unique crashes) and 30.7\% more unique paths (i.e., 5,040 extra unique paths) than the existing  best   seed strategy.

\begin{figure*} 
\begin{center} 
\begin{minipage}[t]{0.195\linewidth}
\centering
\includegraphics[height=2.2cm,width=0.95\textwidth]{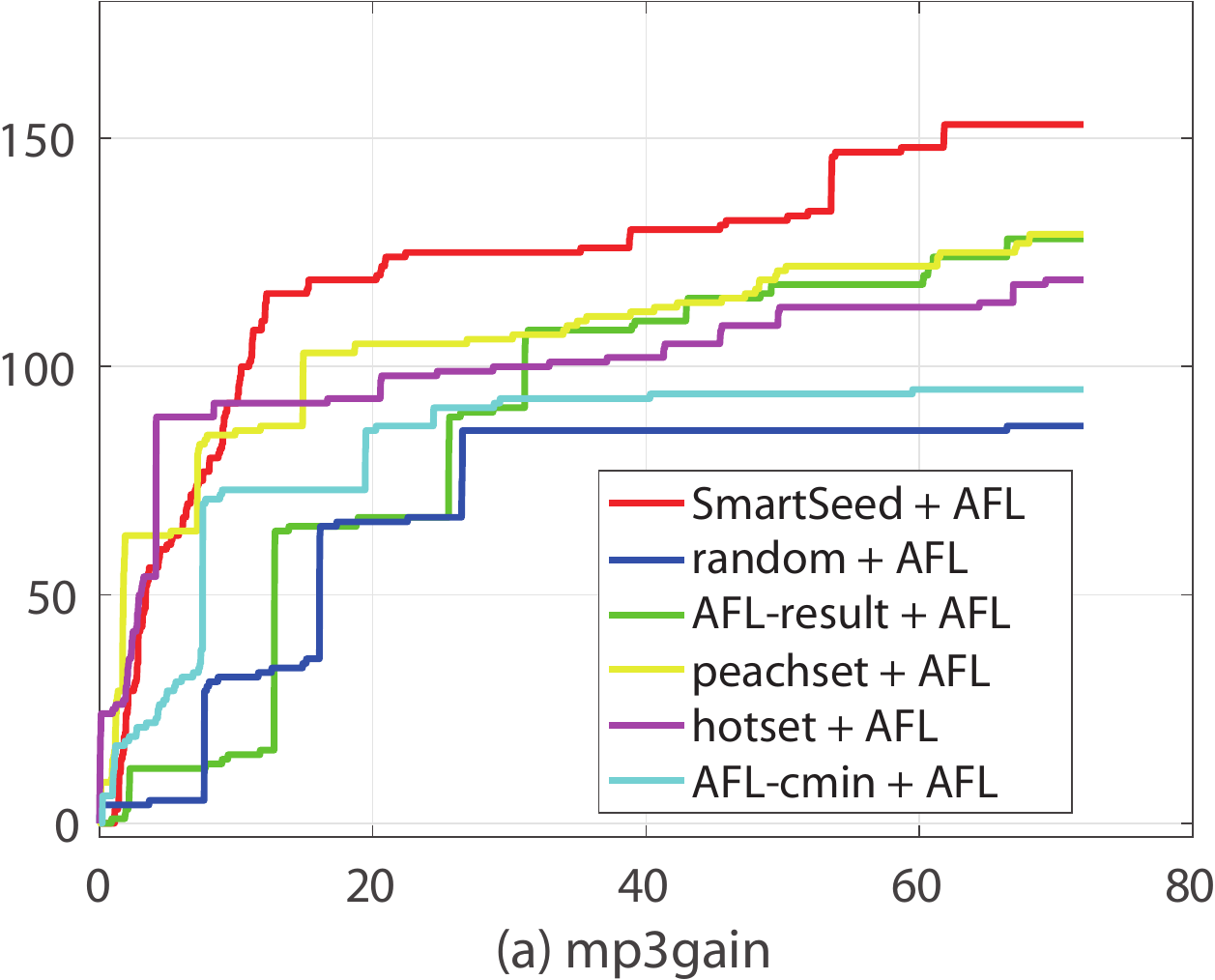}
\end{minipage}
\hfill
\begin{minipage}[t]{0.195\linewidth}
\centering
\includegraphics[height=2.2cm,width=0.95\textwidth]{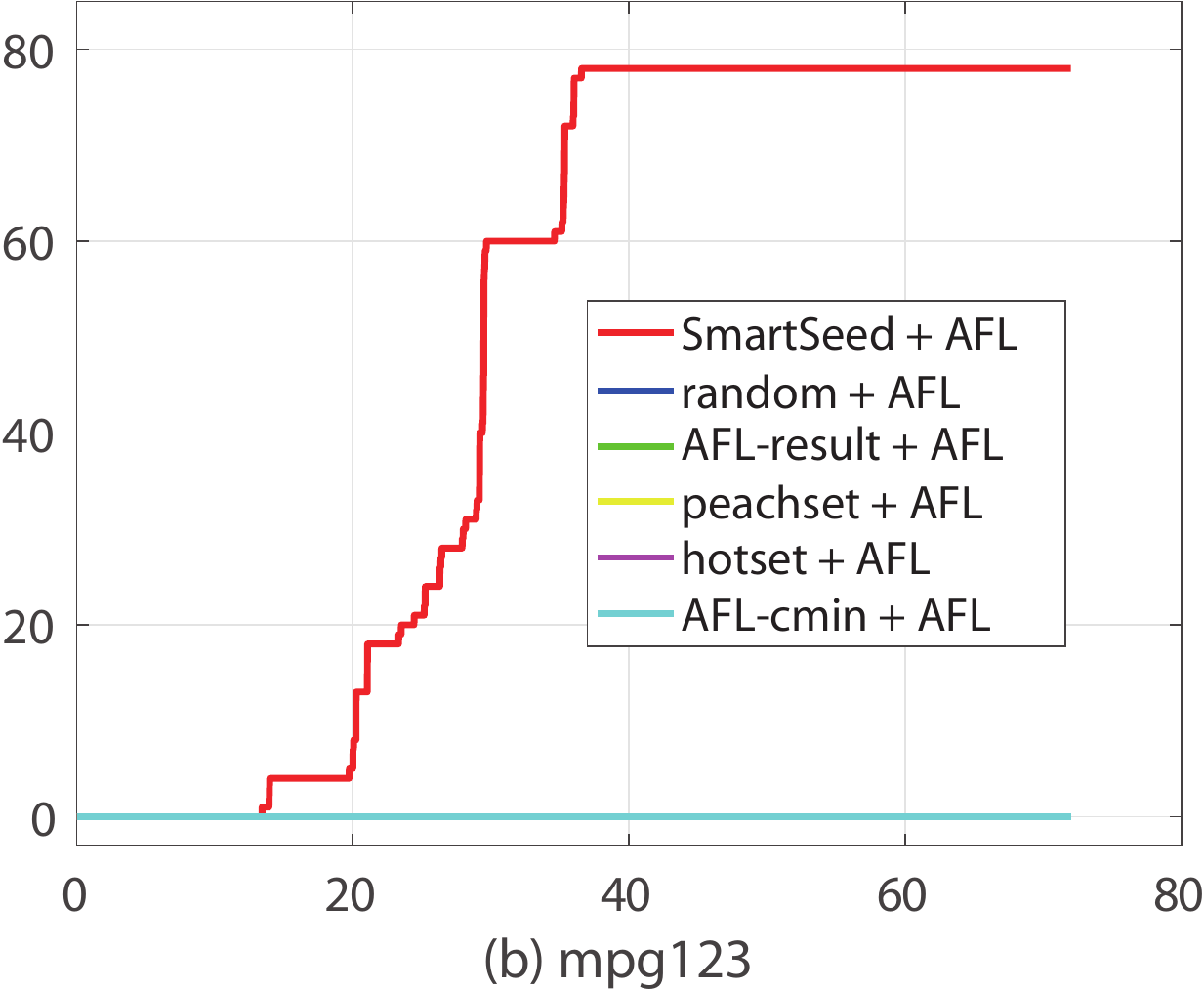}
\end{minipage}%
\hfill
\begin{minipage}[t]{0.195\linewidth}
\centering
\includegraphics[height=2.2cm,width=0.95\textwidth]{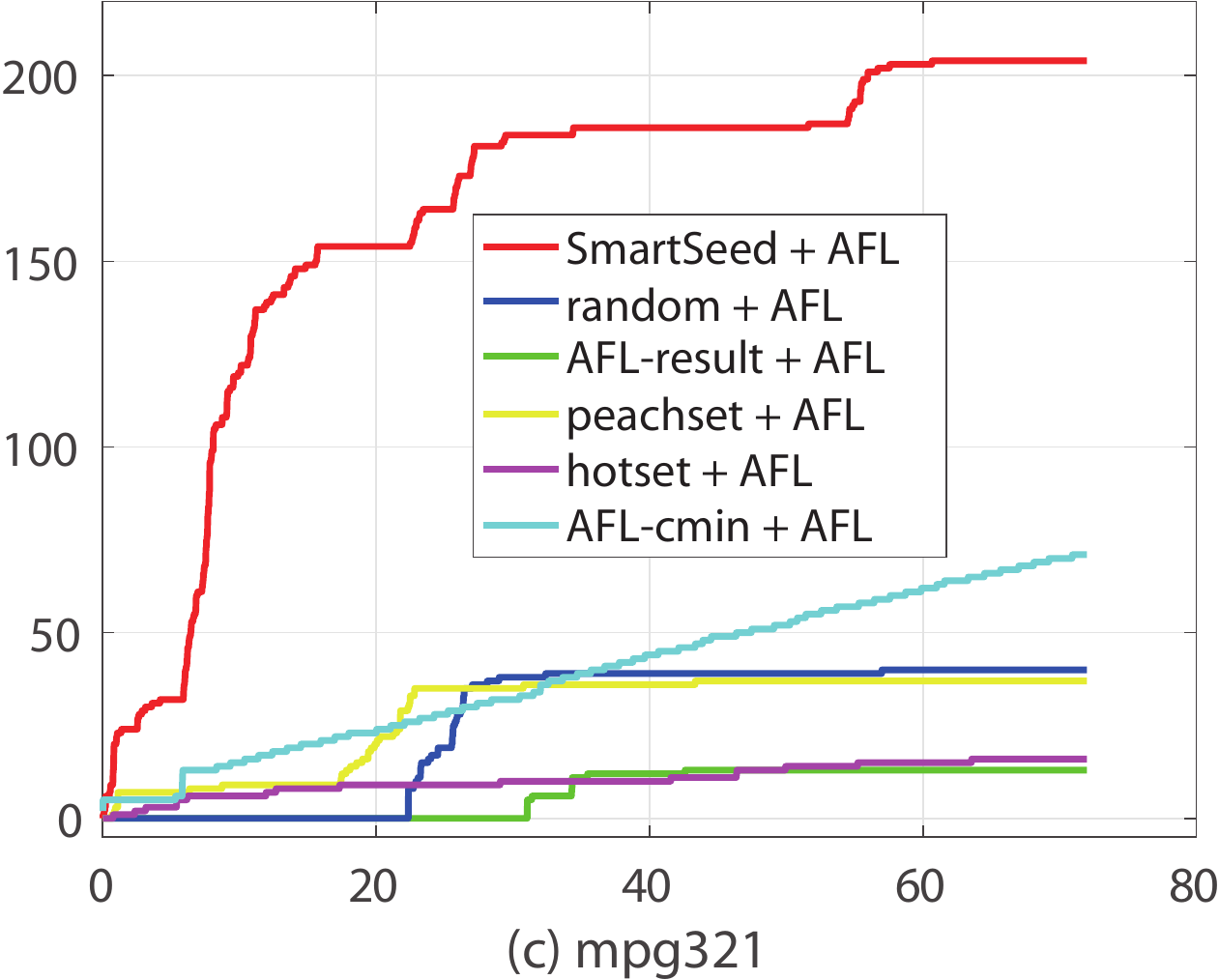}
\end{minipage}
\hfill
\begin{minipage}[t]{0.195\linewidth}
\centering
\includegraphics[height=2.2cm,width=0.95\textwidth]{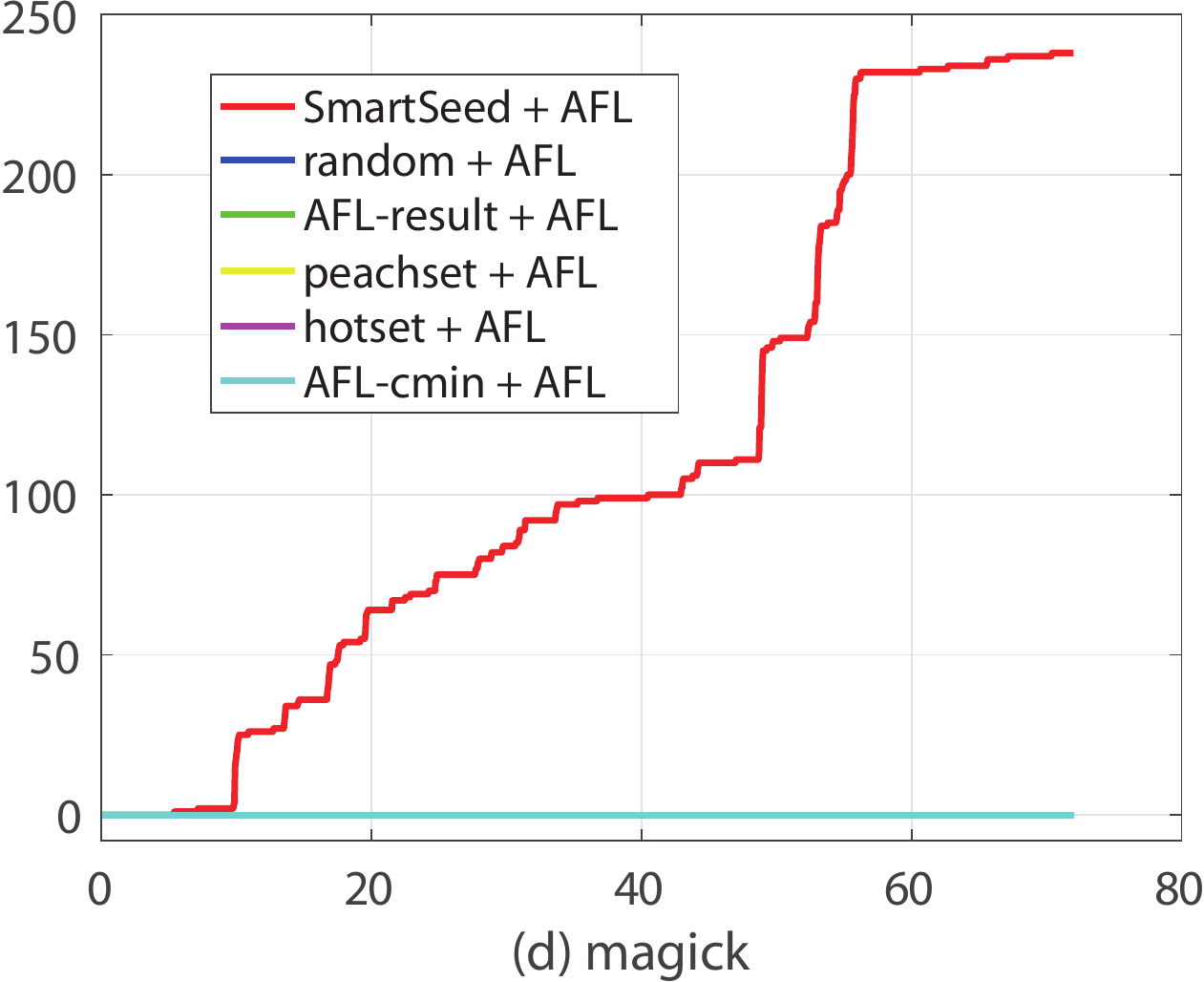}
\end{minipage}
\hfill
\begin{minipage}[t]{0.195\linewidth}
\centering
\includegraphics[height=2.2cm,width=0.95\textwidth]{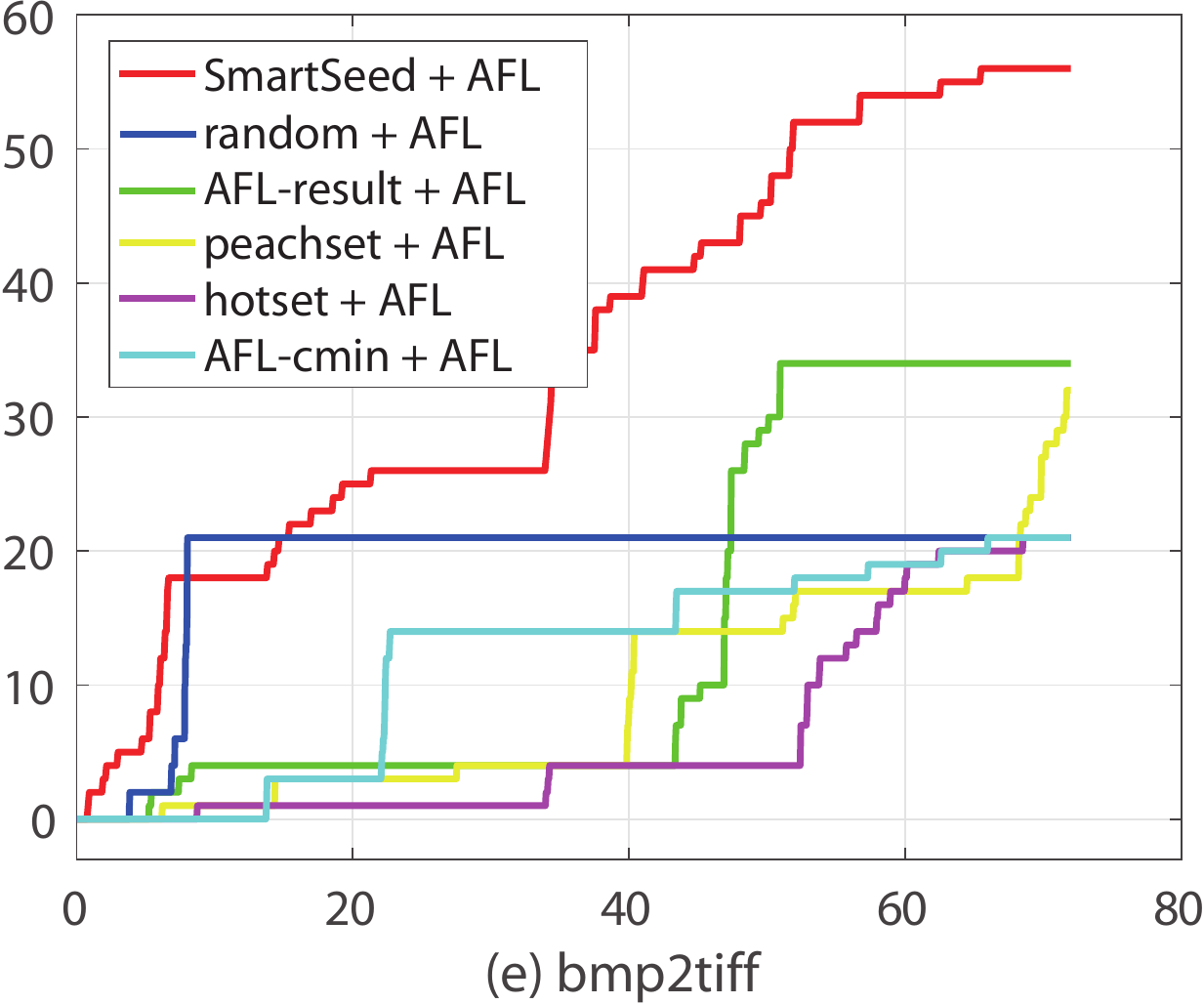}
\end{minipage}
\hfill
\begin{minipage}[t]{0.195\linewidth}
\centering
\includegraphics[height=2.2cm,width=0.95\textwidth]{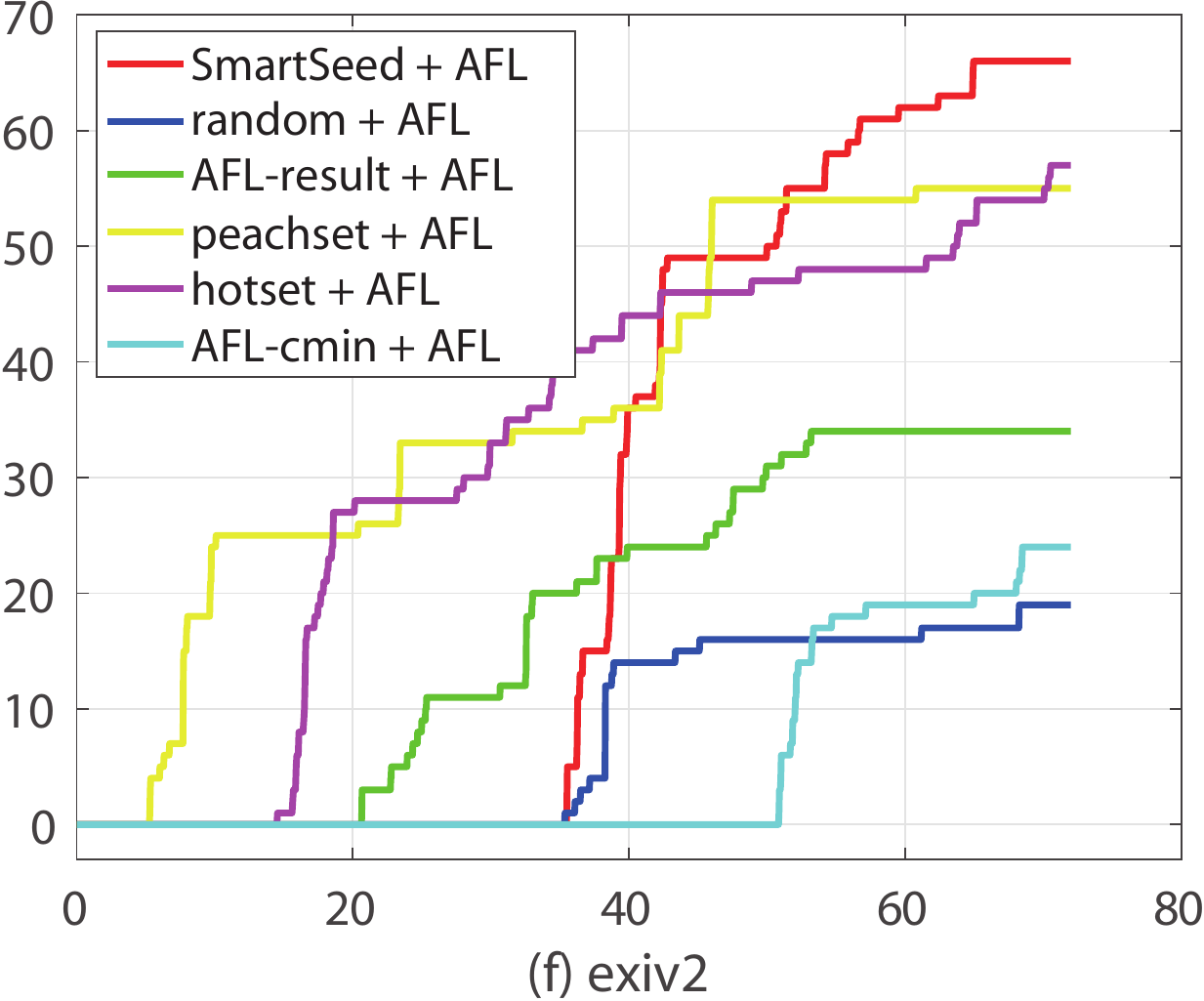}
\end{minipage}
\hfill
\begin{minipage}[t]{0.195\linewidth}
\centering
\includegraphics[height=2.2cm,width=0.95\textwidth]{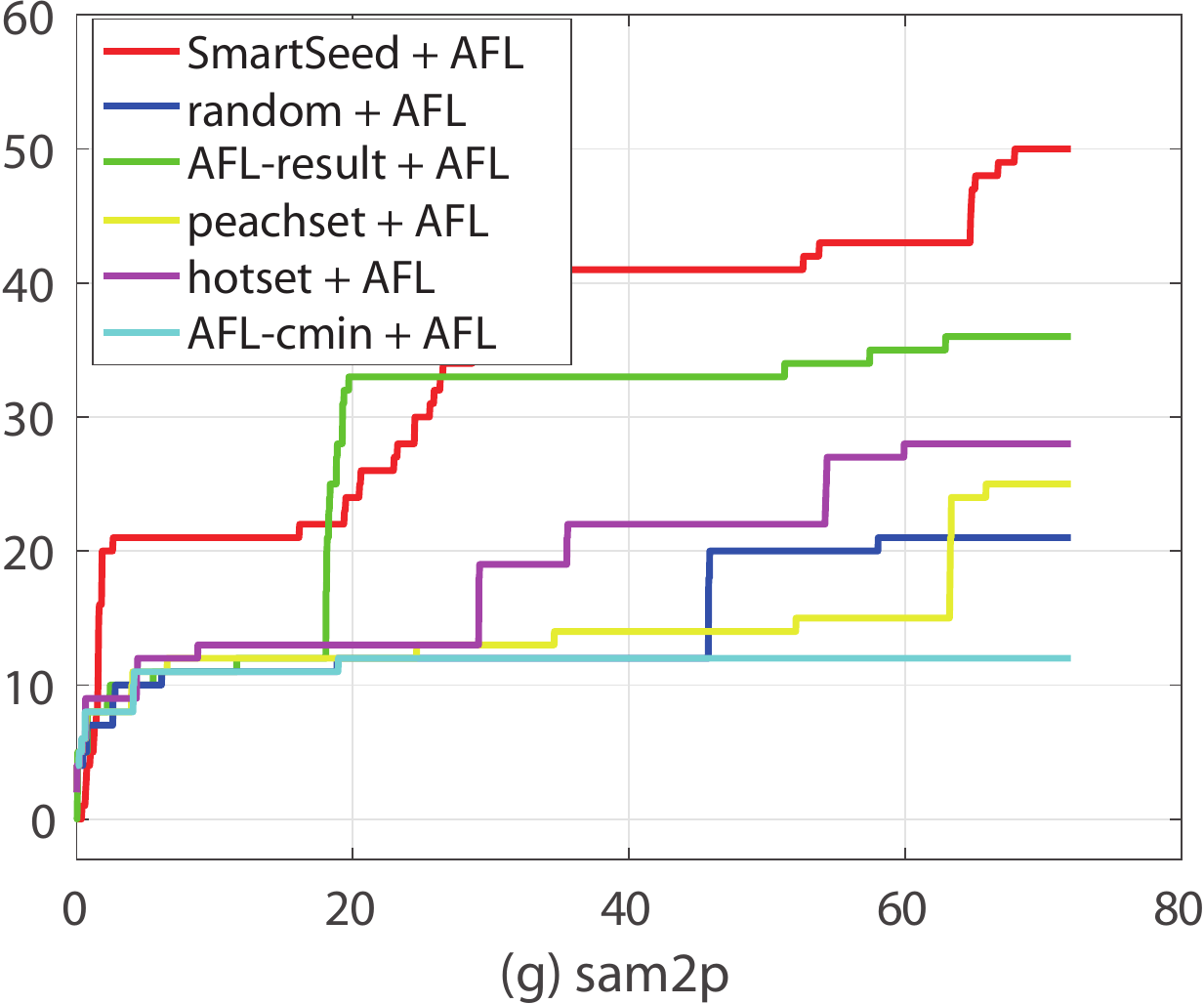}
\end{minipage}
\hfill
\begin{minipage}[t]{0.195\linewidth}
\centering
\includegraphics[height=2.2cm,width=0.95\textwidth]{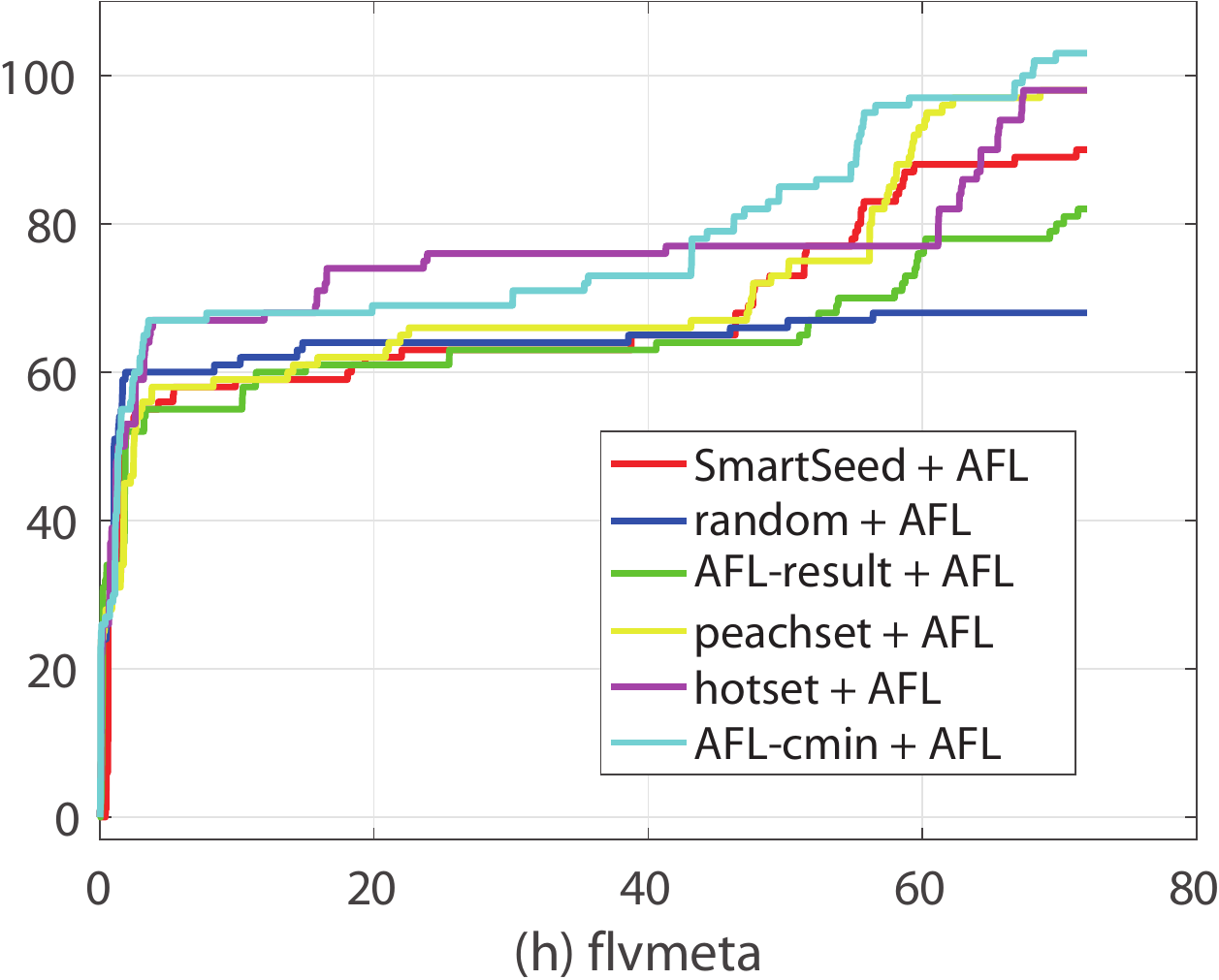}
\end{minipage}
\hfill
\begin{minipage}[t]{0.195\linewidth}
\centering
\includegraphics[height=2.2cm,width=0.95\textwidth]{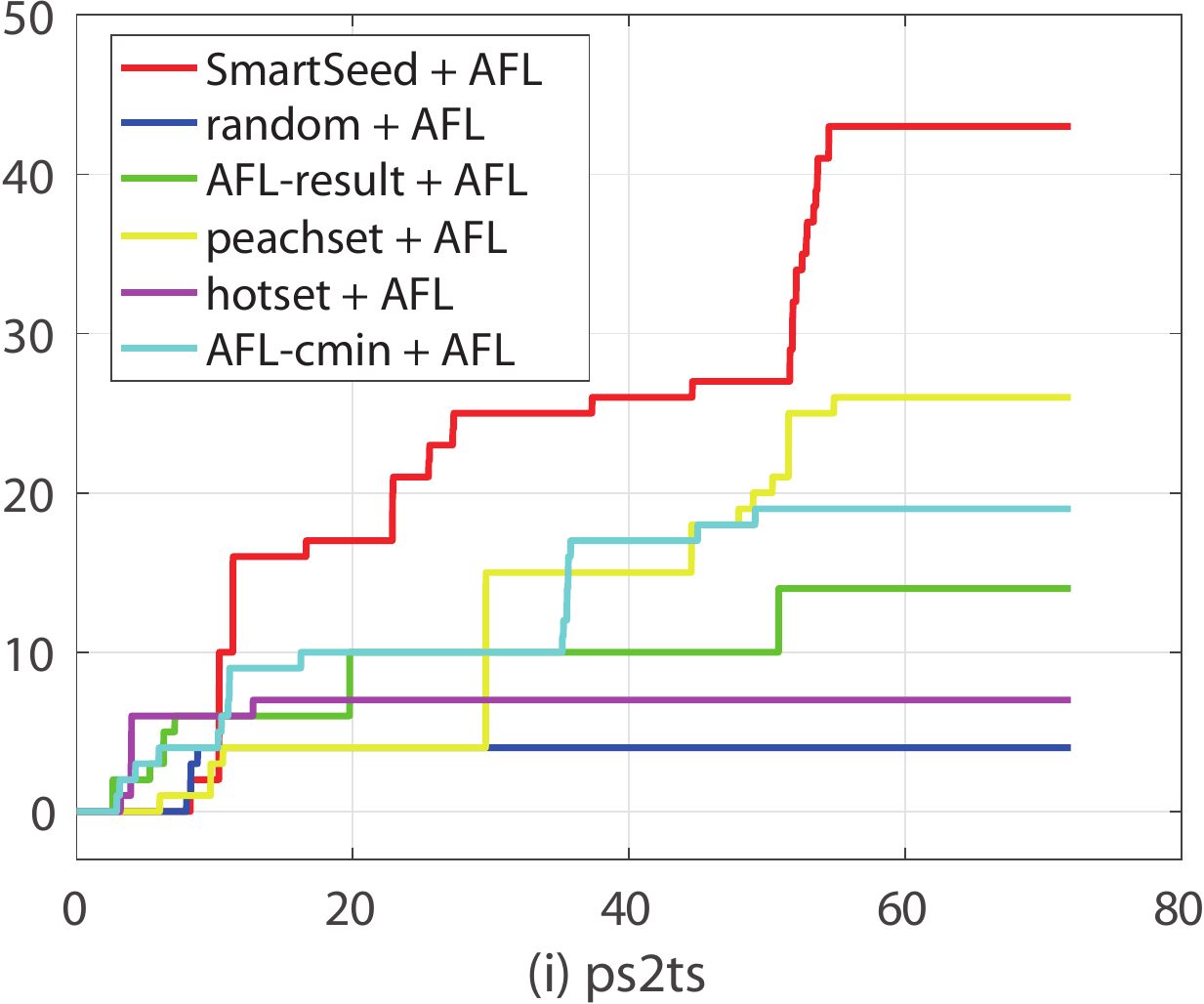}
\end{minipage}%
\hfill
\begin{minipage}[t]{0.195\linewidth}
\centering
\includegraphics[height=2.2cm,width=0.95\textwidth]{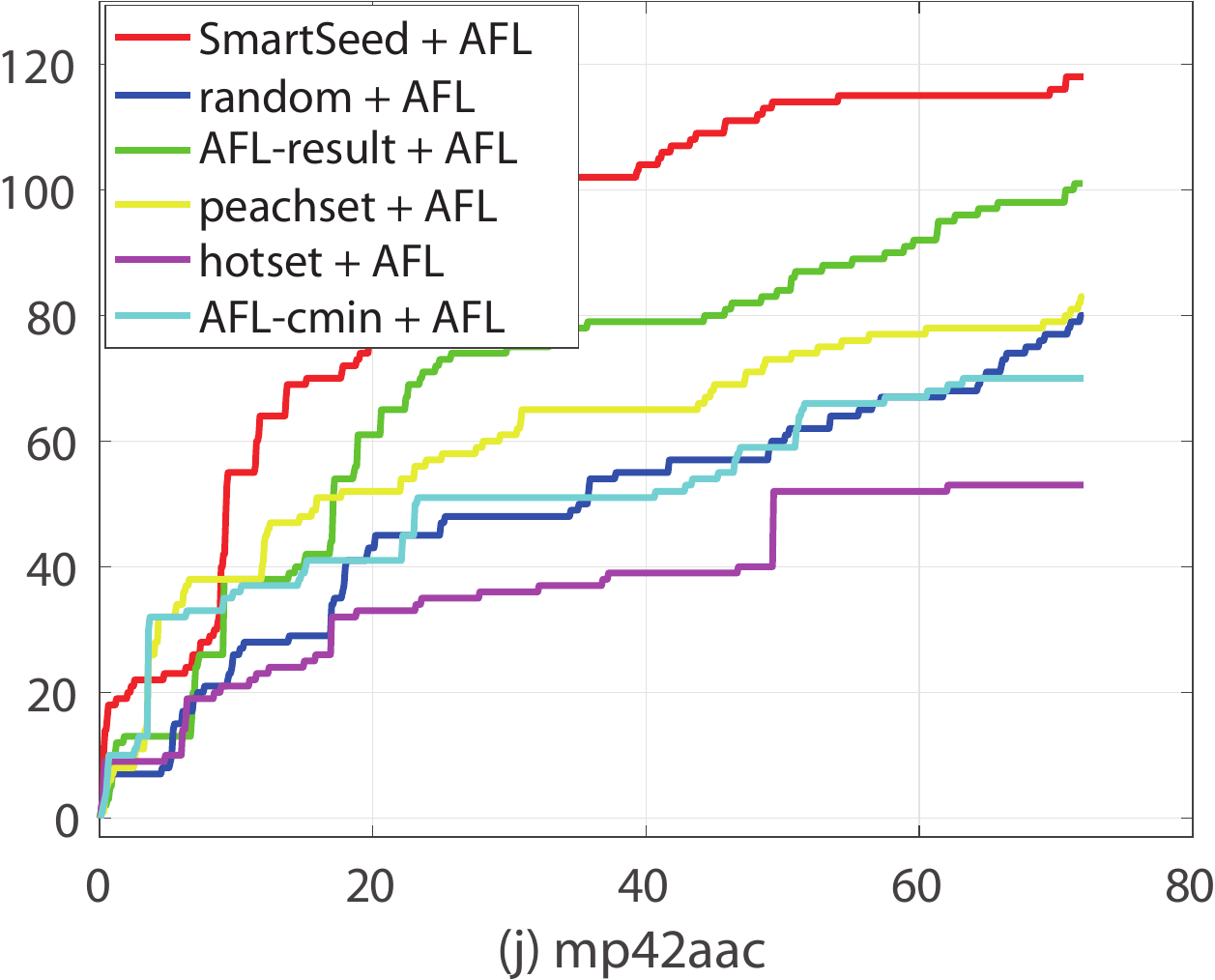}
\end{minipage}
\hfill
\caption{Number of crashes over   72 hours. X-axis: time (over 72 hours). Y-axis: the number of unique crashes.  
}\label{bugtable}

\end{center} 
\end{figure*}

We also visualize the growth of   unique crashes for each   strategy. 
The results are shown in Fig. \ref{bugtable},  from which we have the following observations.

(1) \texttt{SmartSeed} is very efficient, e.g., when fuzzing \texttt{mpg321},   the unique crashes discovered by \texttt{SmartSeed} in 10 hours are more than the crashes   that discovered by other schemes in 72 hours. 

(2) For existing seed selection strategies, they exhibit similar performance during the fuzzing processes and their curves usually mix together. This result is also consistent with the results shown in Table \ref{tableall}.

(3)  
As indicated by the results in Fig. \ref{bugtable},   it takes time for each curve to become stable, which implies  that sufficient time is necessary to enable these strategies to find crashes. This observation is also meaningful for  us to conduct fuzzing evaluation in a proper manner. 

%(1) 
%\texttt{SmartSeed} can spend less time discovering more unique crashes than the results that other fuzzing strategies work  72 hours. For example, It takes 10 hours for \texttt{SmartSeed + AFL} to discover more crashes than the other fuzzing strategies while fuzzing \texttt{mpg321}. 

%(2) 
%The growth of  five state-of-the-art seed selection strategies always mix together, such as while fuzzing  \texttt{mp3gain, bmp2tiff} and \texttt{flvmeta}. This verifies our conclusion that it is hard to say which is the best current seed strategy from the five, they look pretty much the same. 

%(3) 
%In order to evaluate fuzzing strategies, we should run each fuzzing strategy for enough time. If the running time is short, we may obtain the inaccurate results. 
%For example, \texttt{peachset + AFL} was the first one that found the unique crashes,
%while it discovered the third most crashes in 72 hours.

In summary, compared with state-of-the-art seed selection strategies,   
  \texttt{SmartSeed} performs pretty better than other seed strategies for generating   valuable   files  with  multiple input formats. It  not only can discover unique crashes faster, but also can discover more.    

\textbf{Number of Seeds vs. Fuzzing Performance.} To figure out the relationship between the number of seeds and fuzzing performance,  we conduct  the comparative  experiments in the \emph{Supplementary File}  due to the space limitations.    The results show   there is no distinct connection between the number of seeds and fuzzing performance, while it is OK to use 100 seeds as the initial set to guide fuzzing tools in our experiments.

%\subsubsection{Seed Set Size Comparison}
%\emph{}

\subsection{Compatibility}

In this subsection, we examine the compatibility and extendibility of \texttt{SmartSeed}. Specifically, we combine \texttt{SmartSeed} with existing popular fuzzing tools and evaluate its performance.

\subsubsection{Fuzzing Tools} 
In this evaluation, in addition to AFL, we consider the following fuzzing tools. 

%To prove our system can provide the valuable seed set for different fuzzing tools, we combine \texttt{SmartSeed} with the fuzzing tools AFLFast, VUzzer and honggfuzz to fuzz six objective softwares  \cite{item2, item3, item34}. 

%We compare the performances of \textbf{SmartSeed}, \textbf{AFL-result} and \textbf{random} while using these fuzzing tools. The results are shown in Table \ref{tablefuzztool}. 

\texttt{AFLFast \cite{item2}.}    AFLFast is a fuzzing tool based on AFL.  
By using a power schedule to guide the tool towards low-frequency paths, AFLFast  can detect much more  paths with the same execution counts than AFL.

\texttt{honggfuzz \cite{item34}.} honggfuzz is an easy-to-use fuzzing tool provided by \emph{Google}.   Similar to AFL, honggfuzz modifies the input files from the initial seed set and use them for fuzzing.  In   addition, honggfuzz provides powerful process state analysis by leveraging \emph{ptrace}.

\texttt{VUzzer \cite{item3}.} 
VUzzer is a fuzzing tool that focuses on increasing  the coverage. By prioritizing the files mutated from the input files  that reach deep paths, VUzzer can explore more and deeper paths. 
Unlike AFL, AFLFast and honggfuzz who count the edge-coverage, VUzzer uses a dynamic binary  instrumentation tool named  \emph{PIN} to calculate the block coverage, i.e., VUzzer computes the percent of  discovered unique blocks to measure the coverage of an objective application.

\subsubsection{Results}  
For comparison with \texttt{SmartSeed}, we also consider to combine \texttt{random} and \texttt{AFL-result} with the considered fuzzing tools. Specifically, we first employ each seed strategy to obtain 100 seed files,  and then feed each fuzzing tool with these seeds. For the objective applications, we use \texttt{mpg123},  \texttt{mpg321}, \texttt{magick},  \texttt{sam2p}, \texttt{ps2ts} and \texttt{mp42aac}. All the fuzzing  evaluations last for 72 hours and are conducted on same virtual machines with the same settings as in  Section \ref{six}.

	We show the results in Table \ref{tablefuzztool}.   Note that, since honggfuzz failed to build \texttt{sam2p}  and \texttt{mp42aac} by its compiler \texttt{hfuzz-cc}, we cannot count the discovered unique paths for these  two applications when we use honggfuzz to fuzz them. From Table \ref{tablefuzztool}, we have the following  observations.

(1)  With respect to the discovered unique crashes,  
when combining with AFLFast, \texttt{SmartSeed + AFLFast} discovers the most crashes in all the evaluation   scenarios. 
When combining with honggfuzz, \texttt{SmartSeed}  discovers over twice of crashes on \texttt{mpg321} than  the other strategies and is the only one that guides honggfuzz to discover crashes on \texttt{magick}. While on  \texttt{mpg123}, \texttt{sam2p} and \texttt{mp42aac}, all the three seed strategies perform similar.  
When combining with VUzzer, \texttt{SmartSeed} discovers the most crashes on \texttt{mpg321},  \texttt{sam2p}, \texttt{ps2ts} and \texttt{mp42aac}. As for \texttt{mpg123} and \texttt{magick}, no crash is  discovered by the evaluated strategies. All the above results demonstrate that \texttt{SmartSeed} is compatible  with existing popular fuzzing tools, and meanwhile is very effective in fuzzing.

(2)  
With respect to the discovered unique paths (or coverage when combining with VUzzer), \texttt{SmartSeed}  also yields the best performance in most of the evaluation scenarios.    
Specifically, \texttt{SmartSeed} discovers the most unique paths in all the cases when combining with AFLFast.  
As for honggfuzz, except of the error cases (\texttt{sam2p} and \texttt{mp42aac}), \texttt{SmartSeed}  discovers more unique paths than others for \texttt{mpg123} and \texttt{ps2ts}. For \texttt{mpg321}, all the  strategies discover the same number of paths. When combining with VUzzer, we use the criteria as in  \cite{item3}, i.e., we count the block coverage rate instead of unique paths. From the results, all the strategies  discover similar number of paths on mp3 and bmp applications, while \texttt{SmartSeed} yields a much better  performance on flv applications.

%(3)  
%Note that when employing VUzzer for fuzzing \texttt{sam2p}, it gets stuck very frequently. We  conjecture the reason is that VUzzer seems have no countermeasure to deal with the timeout issue of applications.  What's more, when using \texttt{AFL-result + VUzzer} to fuzz \texttt{mpg321}, we   cannot obtain a  convincing result due to its instability and inaccuracy.  %We will discuss  these findings with the authors of VUzzer to seek   solutions for these issues. 

\begin{table*}[] 
	\centering
	\caption{Unique crashes and  paths of each objective application using different fuzzing tools. }\label{tablefuzztool}
	\begin{tabular}{p{1.2cm}<{\centering} p{0.7cm}<{\centering} p{0.8cm}<{\centering} p{0.7cm}<{\centering} p{0.8cm}<{\centering} p{0.7cm}<{\centering} p{0.8cm}<{\centering} p{0.7cm}<{\centering} p{0.8cm}<{\centering} p{0.7cm}<{\centering} p{0.8cm}<{\centering} p{0.7cm}<{\centering}  p{0.8cm}<{\centering}}  
	
		\toprule[1pt]
		\multirow{2}{2.2cm}{ }    & \multicolumn{2}{c}{mpg123} &  \multicolumn{2}{c}{mpg321}          &  \multicolumn{2}{c}{magick}  &  \multicolumn{2}{c}{sam2p} &  \multicolumn{2}{c}{ps2ts}&  \multicolumn{2}{c}{mp42aac}\\ \cline{2-13} 
         & unique crashes & coverage & unique crashes & coverage &  unique crashes & coverage & unique crashes & coverage & unique crashes & coverage & unique crashes & coverage     \\ \midrule[1pt]
         						\multicolumn{1}{l}{SmartSeed + AFL} & \textbf{78} & \textbf{2,154} &\textbf{204}&\textbf{1,060}&\textbf{238}&\textbf{3,374} &\textbf{50}&\textbf{1,322}&\textbf{43}&1,692&\textbf{118}&\textbf{658} \\  
		\multicolumn{1}{l}{random + AFL}    & 0 & 1,183 & 40 & 766&0&697&21&468&4&1,381&80&329 \\
				\multicolumn{1}{l}{AFL-result + AFL} & 0 & 1,001& 13 &187&0&1,149&36&719&14&\textbf{1,740}&102&585 \\ \hline
		\multicolumn{1}{l}{SmartSeed + AFLFast}  &\textbf{148}&\textbf{2,526}&\textbf{161}&\textbf{758}&\textbf{14}&\textbf{2,543}&\textbf{59} &\textbf{1,611} &\textbf{179}&\textbf{2,026} &\textbf{86}&\textbf{592}  \\  
		\multicolumn{1}{l}{random + AFLFast}   &0&1,333 &27&215&1&1,567 &27  & 574&75&1,779&68&569 \\ 
				\multicolumn{1}{l}{AFL-result + AFLFast} &0&1,241&12&194&0&1,090&53 &724 &158&1,787&48&360   \\  \hline
		\multicolumn{1}{l}{SmartSeed + honggfuzz} &0&\textbf{2,776}&\textbf{22}&375&\textbf{3}&5,442&\textbf{10}&error&\textbf{3}&\textbf{1,132}&\textbf{1}&error   \\  
		\multicolumn{1}{l}{random + honggfuzz} &\textbf{1}&534&9&375&0&4,707&9&error&0&1,069 &0&error   \\ 
				\multicolumn{1}{l}{AFL-result + honggfuzz} &0&534&4&375&0&\textbf{5,948}&8&error&\textbf{3}&1,070&\textbf{1}&error   \\  \hline
		\multicolumn{1}{l}{SmartSeed + VUzzer} &0&10\%&\textbf{1,590}&17\%&0&39\%&\textbf{16}&\textbf{19\%}&\textbf{3}&\textbf{14\%}&\textbf{30}&\textbf{8\%}  \\  
		\multicolumn{1}{l}{random + VUzzer}    &0&\textbf{12\%}&1,261&\textbf{18\%}&0&39\%&7&\textbf{19\%}&0&8\%&0&1\% \\
				\multicolumn{1}{l}{AFL-result + VUzzer} &0&\textbf{12\%}& 1,483 & 17\% &0&39\%&2&18\%&0&8\%&0&1\%   \\  
\bottomrule[1pt]
	\end{tabular}
\end{table*}

Overall, \texttt{SmartSeed} exhibits a good compatibility when combing with different fuzzing tools.  Meanwhile, the seeds generated by \texttt{SmartSeed} can significantly improve the performance of existing  popular fuzzing tools.

\subsection{Vulnerability Results}   

To figure out unique vulnerabilities, we recompile the evaluated applications with \texttt{AddressSanitizer} \cite{item60} and use the discovered files of \texttt{SmartSeed}, which triggered   crashes, to test the applications.   
 The results are shown in Table \ref{CVE}, from which we can learn the following  observations. 
%Note that \texttt{AddressSanitizer} runs out of memory when we use files to test \texttt{mp42aac}, thus there is no results for \texttt{mp42aac} in Table \ref{CVE}. 

(1)   Although  we only run each fuzzing process for 72 hours,  from  the crashes  discovered by \texttt{SmartSeed}, we find 23 unique  vulnerabilities in total,  and 16 of them are previously unreported. We submit    them to \emph{CVE} \cite{item59} and have acquired the corresponding CVE IDs. 
This proves that our system is not only  efficient but also can guide fuzzing tools to find more undiscovered  vulnerabilities. 

(2) In total, we discover 9 types of   vulnerabilities. 
This demonstrates that  our system does not  limit on  specific kinds of vulnerabilities,  while \texttt{SmartSeed} can guide fuzzing tools to discover various vulnerabilities. 

%(3) Although  we only run each fuzzing process for 72 hours, \texttt{SmartSeed} discovers 7 CVEs and 24 vulnerabilities that are not previously reported.  This proved that our system is not only efficient but also can guide fuzzing tools to find more undiscovered vulnerabilities. 

\begin{table}[] 
	\centering
	\caption{Vulnerabilities found by \texttt{SmartSeed}.     }\label{CVE}
	\begin{tabular}{p{1cm}<{\centering} p{2.9cm}<{\centering} p{3.23cm}<{\centering} }  
		\toprule[1pt]
	target & type & vulnerability  \\  \midrule[1pt]
  mp3gain &  global-buffer-overflow & CVE-2017-12911; CVE-2017-14410;  CVE-2018-10781(+); CVE-2018-10783(+); CVE-2018-10784(+);      \\
         mp3gain &segmentation violation & CVE-2017-14406;  \\
  mp3gain & stack-buffer-overflow   &  CVE-2018-10777(+); \\
  mp3gain &memcpy-param-overlap & CVE-2018-10782(+); \\
  mpg123 & integer overflow  & CVE-2018-10789(+);\\
  mpg321  & heap-buffer-overflow &  CVE-2018-10786(+);\\
  magick & memory leak & CVE-2017-11754;\\
  bmp2tiff  &heap-buffer-overflow&  CVE-2018-10779(+);\\
  bmp2tiff  &segmentation violation  &  CVE-2014-9330;\\
  exiv2  &heap-buffer-overflow &  CVE-2017-17723;  CVE-2018-10780(+);\\
  exiv2  &stack-overflow &  CVE-2017-14861;\\
        sam2p & heap-buffer-overflow & CVE-2018-10792(+); CVE-2018-10793(+);  \\
   %  sam2p &segmentation violation &  request CVE ID   \\
  %          sam2p  &negative-size-param &  request CVE ID\\
 % flvmeta & heap-buffer-overflow &  request CVE ID \\
    ps2ts &heap-buffer-overflow  & CVE-2018-10787(+); CVE-2018-10788(+); \\
     mp42aac & memory access violation &  CVE-2018-10791(+);  \\
          mp42aac &      buffer overflow &  CVE-2018-10790(+); CVE-2018-10785(+);   \\
 %    mp42aac & buffer over-read  &  request CVE ID\\
  %mp42aac &  & request CVE ID  \\
		\bottomrule[1pt]
	\end{tabular}
\end{table}

\begin{table*}[ht]
	\centering
	\caption{Vulnerabilities found by six fuzzing strategies. }\label{CVETABLE}
	\begin{tabular}{p{0.7cm}<{\centering} p{2.5cm}<{\centering} p{2.5cm}<{\centering} p{2.5cm}<{\centering} p{2.5cm}<{\centering}p{2.5cm}<{\centering} p{2.5cm}<{\centering}  }  
		\toprule[1pt]
	  & SmartSeed + AFL & random + AFL & AFL-result + AFL & peachset + AFL  & hotset + AFL & AFL-cmin + AFL \\  \midrule[1pt]
	  
  mp3gain & CVE-2017-12911; CVE-2017-14406; CVE-2017-14410;  CVE-2018-10781(+); CVE-2018-10783(+); CVE-2018-10784(+); CVE-2018-10782(+);  CVE-2018-10777(+);& CVE-2017-12911; CVE-2017-14410; CVE-2018-10781(+); CVE-2018-10783(+); &CVE-2017-12911; CVE-2017-14410;  CVE-2018-10783(+);  &CVE-2017-12911; CVE-2017-14407; CVE-2017-14410; &CVE-2017-12911; CVE-2017-14407; CVE-2017-14410;   CVE-2018-10777(+); &CVE-2017-12911;  CVE-2017-14407;  CVE-2017-14410;    \\
  
  mpg123 &CVE-2018-10789(+);& & & & &  \\
  
  mpg321  & CVE-2018-10786(+);&CVE-2018-10786(+);&CVE-2018-10786(+);&CVE-2018-10786(+);&CVE-2018-10786(+);&CVE-2018-10786(+); \\ \hline
  
  magick & CVE-2017-11754;& & & & & \\
  
  bmp2tiff  &CVE-2014-9330; CVE-2018-10779(+);& CVE-2014-9330; CVE-2018-10779(+);&CVE-2014-9330;  CVE-2018-10779(+); CVE-2018-10801(+); &CVE-2014-9330; CVE-2018-10779(+);&CVE-2014-9330;  CVE-2018-10779(+);&CVE-2014-9330;  CVE-2018-10779(+);  \\
  
  exiv2  & CVE-2017-14861; CVE-2017-17723;  CVE-2018-10780(+);  & CVE-2017-11339; CVE-2017-14861; CVE-2017-17723;   CVE-2018-9145;&  CVE-2017-11339; CVE-2017-14861;  CVE-2017-14863; CVE-2017-17723; &   CVE-2017-14861;  CVE-2017-17723; &CVE-2017-14861; CVE-2017-17723; &CVE-2017-14861; \\  
  
  sam2p &CVE-2018-10792(+); CVE-2018-10793(+);&CVE-2018-10792(+); &CVE-2018-10792(+); &CVE-2018-10792(+); &CVE-2018-10792(+); &CVE-2018-10792(+); \\ \hline
  
 % flvmeta & new-heap-buffer-overflow-4;&new-heap-buffer-overflow-4;&new-heap-buffer-overflow-4;&new-heap-buffer-overflow-4;&new-heap-buffer-overflow-4;&new-heap-buffer-overflow-4;  \\
  
   ps2ts &CVE-2018-10787(+); CVE-2018-10788(+); &CVE-2018-10787(+); CVE-2018-10788(+); & CVE-2018-10787(+); CVE-2018-10802(+); & CVE-2018-10787(+); CVE-2018-10788(+); & CVE-2018-10787(+); CVE-2018-10788(+); & CVE-2018-10787(+); CVE-2018-10788(+);  \\
   
   mp42aac & CVE-2018-10791(+); CVE-2018-10790(+); CVE-2018-10785(+);&  CVE-2018-10791(+); CVE-2018-10790(+);  CVE-2018-10785(+);&  CVE-2018-10791(+);  CVE-2018-10790(+);  CVE-2018-10785(+); & CVE-2018-10791(+);  CVE-2018-10790(+); &CVE-2018-10791(+);  CVE-2018-10790(+);  &CVE-2018-10791(+);  CVE-2018-10790(+);     \\
  %mp42aac &  & request CVE ID  \\
		\bottomrule[1pt]
	\end{tabular}
\end{table*}

%\textcolor{red}{
In summary, \texttt{SmartSeed} is efficient at discovering various types of       vulnerabilities in practice.    
  Note that recently, the state-of-the-art fuzzer CollAFL \cite{item36} also has fuzzed  \texttt{exiv2}  with the same  version  and   \texttt{mpg123} with a close  version  (ours is 1.25.6, while they used \texttt{mpg123} with version  1.25.0) for  200 hours, and discovered 13 and 1 new vulnerabilities, respectively.   Nevertheless, we  can still discover   new and unreported vulnerabilities on these two  applications using \texttt{SmartSeed} with only 72 hours of fuzzing.  
  This implies that \texttt{SmartSeed} is effective to guide fuzzing tools to find  undiscovered vulnerabilities.

Then, to evaluate the ability of different fuzzing strategies on discovering unique vulnerabilities, we use  input files of all six fuzzing strategies  that trigger crashes to test the applications recompiled by \texttt{AddressSanitizer} \cite{item60}.  
%Since \texttt{AddressSanitizer} fails to locate the line of \texttt{sam2p}'s code triggered by  vulnerabilities, we do not show the vulnerabilities of \texttt{sam2p}.   
Since no fuzzing strategy discovers any crash on \texttt{ffmpeg} or \texttt{avconv} and  the assignment team of \emph{CVE} \cite{item59} dose not regard the crash  on \texttt{flvmeta} as  a vulnerability, we do not  show the results of them. 
%Note that we still make the  vulnerabilities discovered by us anonymous following the submission policy. We add  plus signs after two anonymous vulnerabilities  that are not discovered by \texttt{SmartSeed}.   
The results are shown in Table \ref{CVETABLE},   from which we can learn the following observations. 

(1)  For the already existed CVEs,  all the six fuzzing strategies discover  nearly the same number of  vulnerabilities on \texttt{mp3gain} and \texttt{bmp2tiff}, while they do not find any discovered vulnerability  on \texttt{mpg123}, \texttt{mpg321},  \texttt{sam2p} and \texttt{mp42aac}. Since there is no published CVEs for \texttt{ps2ts} or \texttt{tstools}, no strategy  finds  any   discovered CVE on   \texttt{ps2ts}.   Only \texttt{SmartSeed} finds a discovered vulnerability on \texttt{magick}, while others do not find any crash.  Although  the number of discovered CVEs on \texttt{mp3gain} found by \texttt{random} and \texttt{AFL-result}  is one less than the others,   they perform pretty well on discovering the found CVEs on \texttt{exiv2} and find two more than others. 
As a conclusion, all the six strategies perform  close on discovering the existed CVEs,   with \texttt{SmartSeed}, \texttt{random} and \texttt{AFL-result} find the most, while others find one or two fewer.

(2)   For undiscovered CVEs, \texttt{SmartSeed} finds 16 undiscovered vulnerabilities, which is six more than the second most number discovered by \texttt{random} and \texttt{AFL-result}. However, the number of undiscovered CVEs found by \texttt{peachset}, \texttt{hotset} and \texttt{AFL-cmin} is 7, 8 and 7, respectively. \texttt{SmartSeed} performs pretty well on \texttt{mp3gain}, for which it finds three more undiscovered vulnerabilities  than others.  Only \texttt{AFL-result} finds an undiscovered CVE    on \texttt{bmp2tiff} and another on \texttt{ps2ts} that  are not discovered by  \texttt{SmartSeed}. 

%\textcolor{red}{
In summary, \texttt{SmartSeed} performs the best on finding both discovered and undiscovered vulnerabilities.   
% The results show that \texttt{SmartSeed} finds the most number of   vulnerabilities, which contains   6 more undiscovered CVEs than others.  Note that  unique  vulnerabilities are  pretty rare, we discovered 23 CVEs in total from 1,096 unique crashes found by \texttt{SmartSeed}. Thus, it is a significant improvement that \texttt{SmartSeed} helps AFL discover  6 more undiscovered CVEs.   
In total, we discover 23 CVEs from the 1,096 unique crashes found by \texttt{SmartSeed}  (including 6 more undiscovered CVEs than others).   
 Although \texttt{peachset}, \texttt{hotset} and \texttt{AFL-cmin} require  more time to select seed files  and discover more crashes than the baseline  \texttt{random}, they yield the worst performance on finding unique vulnerabilities,   which is unexpected.

%What's more,  we compare the unique  vulnerabilities discovered by all the six seed   strategies.  The results show that \texttt{SmartSeed} finds the most number of   vulnerabilities, which contains   6 more undiscovered CVEs than others.  Note that  unique  vulnerabilities are  pretty rare, we discovered 23 CVEs in total from 1,096 unique crashes found by \texttt{SmartSeed}. Thus, it is a significant improvement that \texttt{SmartSeed} helps AFL discover  6 more undiscovered CVEs.  
%However,  \texttt{peachset}, \texttt{hotset} and \texttt{AFL-cmin} yield the worst performance.  %Due to the space limitations, more details   are shown in Appendix \ref{APPENDIXCVE}. 

\section{Further Analysis}\label{section9}

To figure out the reasons why \texttt{SmartSeed} performs better than other seed selection strategies, we employ t-SNE \cite{item35}, which is one of the best dimensionality reduction  algorithms that can aggregate similar data together, to visualize the distribution.  
We analyze the similar distribution of the input  seeds generated by  different seed selection strategies  and the mutated files from these seeds, which is a new angle to understand the fuzzing performance  (more details are given in the   \emph{Supplementary File}).  

\subsection{Execution Count} 
For most, if not all, of existing popular coverage-based greybox fuzzing tools like AFL, they are designed to  prioritize  mutating  the seed files that are executed fast. Such design is based on the intuition that a fast-executed seed is more likely to be mutated into input files that are also executed fast, and thus, more tests can be conducted on the objective application within a fixed time, followed by more potential crashes might be found. To measure the execution speed of a seed set (or input files on average), we may use the \emph{execution count}, which is defined as  the number of testing/execution times conducted by the fuzzing tool within a time period.  Evidently, a larger execution count implies more input files can be mutated  from the seeds to test the application  and faster execution speed on average for each input file. 
%In regard to execution count, the fuzzing tools, such as AFL, use the execution time of an input file as the one of  indicators to determine how many new input files are generated from this input file. In other words, these fuzzing tools are more willing to generate   input files whose execution speed are faster. 
%However, we can learn the antithetic results from Table \ref{tableall} and Table \ref{count}. 

Now, under the same settings as the experiments in Table \ref{tableall}, we analyze the execution count of the seeds generated by \texttt{SmartSeed} and state-of-the-art seed selection strategies. The results are shown in Table \ref{count}. Considering the results in Tables \ref{tableall} and \ref{count} together, we find that:  although \texttt{SmartSeed + AFL} is the most effective strategy in most of the evaluation scenarios with respect to discovering both crashes and paths, it does not have the largest execution count in most of the cases, i.e., \texttt{SmartSeed + AFL} discovers more crashes and paths using fewer input files. For instance, \texttt{SmartSeed + AFL} is the only strategy that discovers crashes on \texttt{magick} while its execution count is the least one. Therefore, based on the results in Tables \ref{tableall} and \ref{count}, we conclude that \emph{there is no explicit correlation between the execution count and the number of crashes and paths being discovered}.

\begin{table*}[] 
	\centering
	\caption{Execution count and generation of each objective application under different fuzzing strategies. }\label{count}
	\begin{tabular}{p{1.5cm}<{\centering} p{0.9cm}<{\centering} p{0.9cm}<{\centering} p{0.9cm}<{\centering} p{0.9cm}<{\centering} p{0.9cm}<{\centering} p{0.9cm}<{\centering} p{0.9cm}<{\centering} p{0.9cm}<{\centering} p{0.9cm}<{\centering} p{1.1cm}<{\centering} p{0.9cm}<{\centering}  p{0.9cm}<{\centering}}

		\toprule[1pt]
		\multirow{2}{2.2cm}{execution count            /generation}    & \multicolumn{4}{c}{mp3} &  \multicolumn{4}{c}{bmp}          &  \multicolumn{4}{c}{flv}   \\ \cline{2-13} 
		                                                                  & mp3gain & ffmpeg & mpg123 & mpg321 &  magick & bmp2tiff & exiv2 & sam2p & avconv & flvmeta &ps2ts & mp42aac     \\ \midrule[1pt]
		\multicolumn{1}{l}{Smartseed + AFL} & 39.9M/\textbf{15} & 12M/2  & 49.5M/\textbf{26} & 50.5M/22  & 6.5M/\textbf{11}   & 3.8M/\textbf{8}   &45.9M/14   & 4.5M/\textbf{8}   & \textbf{20.7M}/3   & 498.1M/27 & 340M/\textbf{23}  & \textbf{26.7M}/\textbf{21}   \\ 
				\multicolumn{1}{l}{random + AFL}          & 34.5M/3 & \textbf{22.9M}/\textbf{3}  & 56.1M/3 & 40.1M/\textbf{26}    & \textbf{24.8M}/3   & 41.5M/3  & 57.5M/14  & 33.9M/3   & 20.5M/3   & 338.9M/17 & 231M/21 & 5.2M/17  \\ 
		\multicolumn{1}{l}{AFL-result + AFL} & 32M/3 & 14.7M/\textbf{3}  & 58.5M/3 &37.7M/10   & 15.7M/2  & 25.8M/4   & \textbf{73.2M}/\textbf{22}   & \textbf{74.3M}/3   & 11.7M/3   & 454.7M/27 & \textbf{366M}/21  & 9.4M/18     \\ 
		\multicolumn{1}{l}{peachset + AFL}    & 55.7M/3 &8.9M/2& 45.2M/3&47.1M/13&21.7M/3&37.5M/4&27.7M/16&44.5M/3&13.3M/3&441.7M/27& 228M/22 &10.9M/19 \\ 
		\multicolumn{1}{l}{hotset + AFL}    &\textbf{59.9M}/3&9.4M/2&\textbf{80.2M}/3&42.4M/4&24.1M/4&32.2M/4&34.8M/18&56.3M/3&11.8M/3&421.3M/25&152M/20&1.9M/13 \\   
		\multicolumn{1}{l}{AFL-cmin + AFL}    &38.4M/3 &9.7M/2&66.9M/3&\textbf{61.8M}/4&16.8M/3 &\textbf{48.5M}/3 &61.3M/9&41.9M/2&17.8M/3&\textbf{664.3M}/\textbf{29}&343M/\textbf{23}&3.1M/13   \\
		\bottomrule[1pt]
	\end{tabular}
\end{table*}

The above analysis leads to an interesting insight. \emph{It might be a misunderstanding that seed files with  faster execution speed can discover more unique crashes or paths}. Although these seeds can generate more files  and test the objective application more times, most of generated files may execute the known paths and thus may  discover nothing. Therefore, \emph{valuable seeds are more desired for efficient fuzzing rather than fast-executed ones}. 

%These results lead to an interesting finding: \textbf{ It is a misunderstanding that the seed files with the faster execution speed can discover more unique crashes during the same time. Although generating the input files with the faster execution speed can test the objective software more times, most of them execute the known paths and discover nothing. 
%What the fuzzing tools need are the input files that are easier to discover the crashes, rather than  execute fast. }

%Since it is difficult for the fuzzing tools to speculate which input files certainly can trigger the crashes, \texttt{SmartSeed} is valuable as it can perform the best while fuzzing the most objective softwares. 

\subsection{Generation Analysis} 
%In regard to generations, we  can learn the following phenomena from Table \ref{tableall} and Table \ref{count}. 
For genetic algorithm based fuzzing, we usually use \emph{generation} to measure the mutation/generation-relationship between an input file and the seed files. For instance, the generation of the initial seed files is ``1'',  the files that are mutated/generated from the seed files have a generation of ``2'', and similarly, the files that are  mutated/generated from the files with generation \emph{k} have a generation of \emph{k}+1.

%Then, under the same settings as the experiments in Table \ref{tableall}, we show the largest generation number among all the generated files of each seed strategy in Table \ref{count}. Considering the results in Tables \ref{tableall} and \ref{count} together, we find that in general, a large achieved generation implies better fuzzing performance. For instance, the largest achieved generation of \texttt{SmartSeed} is much larger than other strategies on \texttt{mp3gain, mpg123, magick, bmp2tiff, sam2p} and \texttt{mp42aac}, and meanwhile, \texttt{SmartSeed} has better fuzzing performance on these softwares. This is also why the valuable mutated input files of \texttt{SmartSeed} spread in a larger area as shown in Fig. \ref{crashdistribution}. 

Then, under the same settings as the experiments in Table \ref{tableall}, we show the largest generation number  among all the generated files of each seed strategy in Table \ref{count}. Considering the results in Tables  \ref{tableall} and \ref{count} together, we find that in general, a large achieved generation implies a better  coverage performance.  
%This is also why the valuable mutated input files of \texttt{SmartSeed} spread in a larger area as shown in Fig. \ref{crashdistribution}. 
In 11 of the evaluated 12 applications except for \texttt{avconv},  the most unique paths are discovered by the fuzzing     strategy that has the largest or the second largest generation. 	 
For instance, the largest achieved generation of \texttt{SmartSeed} is much larger than   other seed generation/selection strategies on \texttt{mp3gain}, \texttt{mpg123}, \texttt{magick}, \texttt{bmp2tiff}, \texttt{sam2p} and \texttt{mp42aac}, and meanwhile, \texttt{SmartSeed} has a  better coverage performance on these applications.

%(1) 
%When the generations is large, the number of discovered unique paths  is large. For example, only the generations of \texttt{SmartSeed + AFL} is 15 while fuzzing \texttt{mp3gain}. Then our fuzzing strategy discovered more than  750 paths than the other  strategies. 
%The similar phenomena  appeared while fuzzing \texttt{mpg123, magick, exiv2, sam2p} and \texttt{mp42aac}. 
%Thus, we get another interesting finding: \textbf{ The larger the generations is, It is more likely for the fuzzing strategy  to  discover more  unique paths. }

%(2) 
%We figure out the  statistical data from Table \ref{count}. \texttt{SmartSeed + AFL} has the largest generations while fuzzing the 8 of all 12 objective softwares for 72 hours (\texttt{mp3gain, mpg123, magick, bmp2tiff, sam2p, avconv, ps2ts, mp42aac}). Based on the interesting finding above, \texttt{SmartSeed} is likely to generate the seed files that discover more unique paths. 

%It is worth to note that  many researchers do a lot of work to  increase the coverage in order to find more crashes. Now we learn the fact that a better seed set also can improve the coverage. 

Therefore, based on our results and analysis, we have another interesting insight: it is very likely that   \emph{there is a positive correlation between the  largest achieved generation and the coverage performance of seed   strategies}.  This can explain the significant coverage improvement of \texttt{SmartSeed}.  

Note that many researchers do a lot of work to  
increase the coverage in order to find more crashes. Now we learn  
the fact that a better seed set  can  also improve the coverage.

%In summary, our results demonstrate that  execution count  is not the appropriate indicator to evaluate the value of an input file. 
%What's more, the results show  that the larger generations leads to more unique paths, 
%which explains the reason why \texttt{SmartSeed} is easier to discover more unique paths. 

\section{Discussion}\label{section5}

In this section, we make some discussions on \texttt{SmartSeed} starting from the three proposed heuristic questions in Section \ref{section1}.  Then, we remark the limitation and future work of this paper. 

\textbf{Q1: Can we generate valuable seeds in a fast and effective manner?}   
From Table \ref{tabletime}, the seed generation process of \texttt{SmartSeed} is linearly scalable and   fast. It only takes tens of seconds to generate sufficient seed files for  popular fuzzing tools.   Furthermore, as we evaluated in Section \ref{section4}, \texttt{SmartSeed}  significantly outperforms existing  state-of-the-art seed selection strategies when fuzzing multiple kinds of  practical applications. Therefore, as  expected,  \texttt{SmartSeed} can generate effective seeds in a fast and effective manner.

\textbf{Q2: Can we generate valuable seeds in a robust manner?}      
As shown in Section \ref{section3}, \texttt{SmartSeed} is designed as a generic system to generate valuable seeds for applications without of requiring highly-structured formats. We also implement this system to generate seeds for applications with different formats. As shown in Section \ref{section4}, once the generative model of \texttt{SmartSeed} is constructed, it can be employed to fuzz many applications with the same/similar format, and meanwhile, its performance significantly outperforms state-of-the-art seed selection techniques. Therefore, based on our evaluation, \texttt{SmartSeed} can generate valuable seeds in a robust manner.

\textbf{Q3: Can we generate valuable seeds in a compatible manner?}   
As shown in Section \ref{section4}, \texttt{SmartSeed} is easily compatible with existing popular fuzzing tools. Furthermore, their fuzzing performance can also be improved in most of the scenarios. Therefore,  \texttt{SmartSeed} is easily extendable and compatible.

% Limitations and future works of our paper are shown in  the \emph{Supplementary File}  due to the space limitations. 

Due to the space limitations, we show the limitation and future work of this paper in the \emph{Supplementary File}.

\section{Conclusion}\label{section7}
In this paper, we present a novel unsupervised learning system named \emph{SmartSeed} to generate valuable input seed files for   fuzzing.   
%The main procedures of SmartSeed are as follows: (1) collect the valuable training set by using fuzzing; (2) convert files in the training set to matrices; (3) use matrices to train WGAN model; (4) generate similar matrices by the generative model of WGAN model and convert them to the files as input set. 
%Through the experiments we proved that SmartSeed can not only efficiently generate multiple input formats, but also have the better performance of the fuzzing tools than the files from AFL or the Internet in most cases. 
Compared with   state-of-the-art seed selection strategies, \texttt{SmartSeed} discovers   608 extra  unique crashes  and 5,040 extra unique paths when we use AFL to fuzz 12 open source applications with different input formats.   
%Only \texttt{SmartSeed + AFL} discovered bus error on \texttt{mpg123} and   segmentation fault on \texttt{magick}. 
Then, we combine  \texttt{SmartSeed} with different fuzzing tools. The evaluation results demonstrate that \texttt{SmartSeed} is easily compatible and meanwhile very effective.  
To further understand  the performance of \texttt{SmartSeed},  
%we analyzed the distributions of input sets and bug triggers. 
%Next, we analyzed the execution  count  and generations of all six seed   strategies. 
we make more analysis on the seeds generated by \texttt{SmartSeed} and present  
  several interesting findings to enlighten our knowledge on effective fuzzing.  
What's more, we   find  23 unique vulnerabilities on 9 applications by \texttt{SmartSeed}  and have  obtained CVE IDs for 16 undiscovered ones. 
Finally,  %we published the prototype of \texttt{SmartSeed}. 
we will open source the \texttt{SmartSeed}  system,  which is expected to facilitate future fuzzing research.

\newpage
\emph{ }
\newpage
\section*{\textbf{Supplementary File of ``SmartSeed: Smart Seed Generation for Efficient Fuzzing''}}
%
%
% author names and IEEE memberships
% note positions of commas and nonbreaking spaces ( ~ ) LaTeX will not break
% a structure at a ~ so this keeps an author's name from being broken across
% two lines.
% use \thanks{} to gain access to the first footnote area
% a separate \thanks must be used for each paragraph as LaTeX2e's \thanks
% was not built to handle multiple paragraphs
%

% For peer review papers, you can put extra information on the cover
% page as needed:
% \ifCLASSOPTIONpeerreview
% \begin{center} \bfseries EDICS Category: 3-BBND \end{center}
% \fi
%
% For peerreview papers, this IEEEtran command inserts a page break and
% creates the second title. It will be ignored for other modes.
\IEEEpeerreviewmaketitle

\section{Number of Seeds vs. Fuzzing Performance}   \label{NumberofSeeds}

\begin{table*}[ht]\footnotesize
	\centering
	\caption{Unique crashes and  paths of each objective application when using different number of files as the initial seed set to fuzz. }\label{differentnum}
	\begin{tabular}{p{1.2cm}<{\centering} p{1.2cm}<{\centering} p{2cm}<{\centering} p{2cm}<{\centering} p{2cm}<{\centering} p{2cm}<{\centering} p{2cm}<{\centering}  p{2cm}<{\centering} p{2cm}<{\centering}}  
		\toprule[1pt]
	 &	\multirow{2}{2.2cm}{ number }    & \multicolumn{2}{c}{SmartSeed + AFL} &  \multicolumn{2}{c}{random + AFL}          &  \multicolumn{2}{c}{AFL-result + AFL}  \\ \cline{3-8} 
     &  & unique crashes & unique paths & unique crashes & unique paths &  unique crashes & unique paths       \\ \midrule[1pt]
		\multirow{4}{8cm}{mpg321 } 
		& 50 &\textbf{231}&1,029&48&\textbf{784}&\textbf{25}&\textbf{329}	      \\   
			  & 100 &204&\textbf{1,060}&40&766&13&187     \\   
			  &200&164&1,051&\textbf{66}&565&10&126  \\   
		  &	300 &181&1,023&24&608&9&104   \\  \hline
		\multirow{4}{8cm}{sam2p } 
		&	50&37&1,358&23&\textbf{559}&21&404     \\   
			  & 100&\textbf{50}&1,322&21&468&\textbf{36}&\textbf{719}      \\  
			  & 200 &49&\textbf{1,392}&27&458&30&372   \\   
		  &	300 &48&1,373&\textbf{28}&494&34&357    \\  \hline
		  \multirow{4}{8cm}{mp42aac } 
		  &	50&\textbf{176}&\textbf{1,226}&76&560&75&441      \\   
			  & 100&118&658&80&329&\textbf{102}&\textbf{585}    \\  
			  &  200&108&879&83&\textbf{626}&31&340  \\   
		  &	300 &106&502&\textbf{91}&520&53&473   \\  %\hline  
%		\multicolumn{1} {l} {total}  &&  \\   
%		\multicolumn{1} {l} {average} &&  \\
\bottomrule[1pt]
	\end{tabular}
\end{table*}

Now, we examine the relationship between the number of seed files and the fuzzing effectiveness.  
We use the same  virtual  machines  with  an Intel i7 CPU,  4.5GB memory   and a Ubuntu 16.04 LTS system    to run each fuzzing process for  72 hours. 
The results are shown in Table \ref{differentnum}, from which we can learn the following observations. 

(1) Because mutation-based fuzzing tools such as AFL use byte flipping and so on to mutate files, the variation of the mutated files is slight and slow. Therefore, the fuzzing performance will be  influenced if the number of seeds is too small due to the lack of   seeds' diversity. 
%Thus, we have to test the lower limit of the number of seeds. 
Then, a proper number of seeds should be bigger than the lower bound.   
Since the performance  of all  the three fuzzing strategies are not obviously getting worse  when the number of seeds in the initial set is 50, we come to a conclusion  that 50 or more   seeds is enough  to bootstrap these   fuzzing tools.  

(2) In regard to the discovery of unique crashes,  %\texttt{SmartSeed + AFL} discovers much more crashes than others.    
for \texttt{mpg321}, \texttt{SmartSeed} and \texttt{AFL-result} perform the best when the number of seeds   is 50, while \texttt{random} discovers the most crashes when the seed number is 200. 
\texttt{SmartSeed} and \texttt{AFL-result} discover the most crashes on \texttt{sam2p} when the seed number is 100, while \texttt{random} finds the most unique crashes on  \texttt{sam2p}  with 300 seeds.  
For \texttt{mp42aac},  \texttt{SmartSeed} finds the most crashes when the seed number is 50.    \texttt{random} and \texttt{AFL-result} discover the most crashes when the number of seeds is 300 and 100, respectively.  
Therefore, it seems  that there is no obvious relationship between the number of discovered crashes and the number of seeds   when there are enough seeds.

(3) As for the discovery of unique paths,   
for \texttt{mpg321}, \texttt{SmartSeed} finds the most paths when the number of seeds is 100, while \texttt{random} and \texttt{AFL-result} discover the most paths when they have 50 seeds.   
When fuzzing \texttt{sam2p}, \texttt{SmartSeed} finds the most paths with 200 seeds, while \texttt{random} uses 50 seeds to discover the most paths. \texttt{AFL-result} finds the most paths when the number of seeds is 100. 
For \texttt{mp42aac},  \texttt{SmartSeed}, \texttt{random} and \texttt{AFL-result} find the most paths  when the number of  seeds  is 50, 200 and 100, respectively. 
It seems that if the number of the initial seeds is too big, the number of discovered paths will  decrease rarely.  
We  conjecture the reason is that the initial seeds detect more unique paths, and thus, the discovered paths later will be fewer. 
Therefore, it is a negligible  relationship between the number of seeds in the initial set and the number of discovered paths. 
 Note that it is hard to say 100 is a big number for the number of seeds in the initial set.

In summary, it seems 50 seeds in the initial seed set  is enough to guide the fuzzing tools to detect crashes and paths, while there is no evident relationship between the number of seeds and the fuzzing performance.  Therefore,  it is OK to use 100 seeds as the initial seed set  in the main body of our paper.

\section{Further Analysis}\label{section9}

\begin{figure*} [tp]
\begin{center} 
\begin{minipage}[t]{0.33\linewidth}
\flushright
\includegraphics[height=3cm,width=0.8\textwidth]{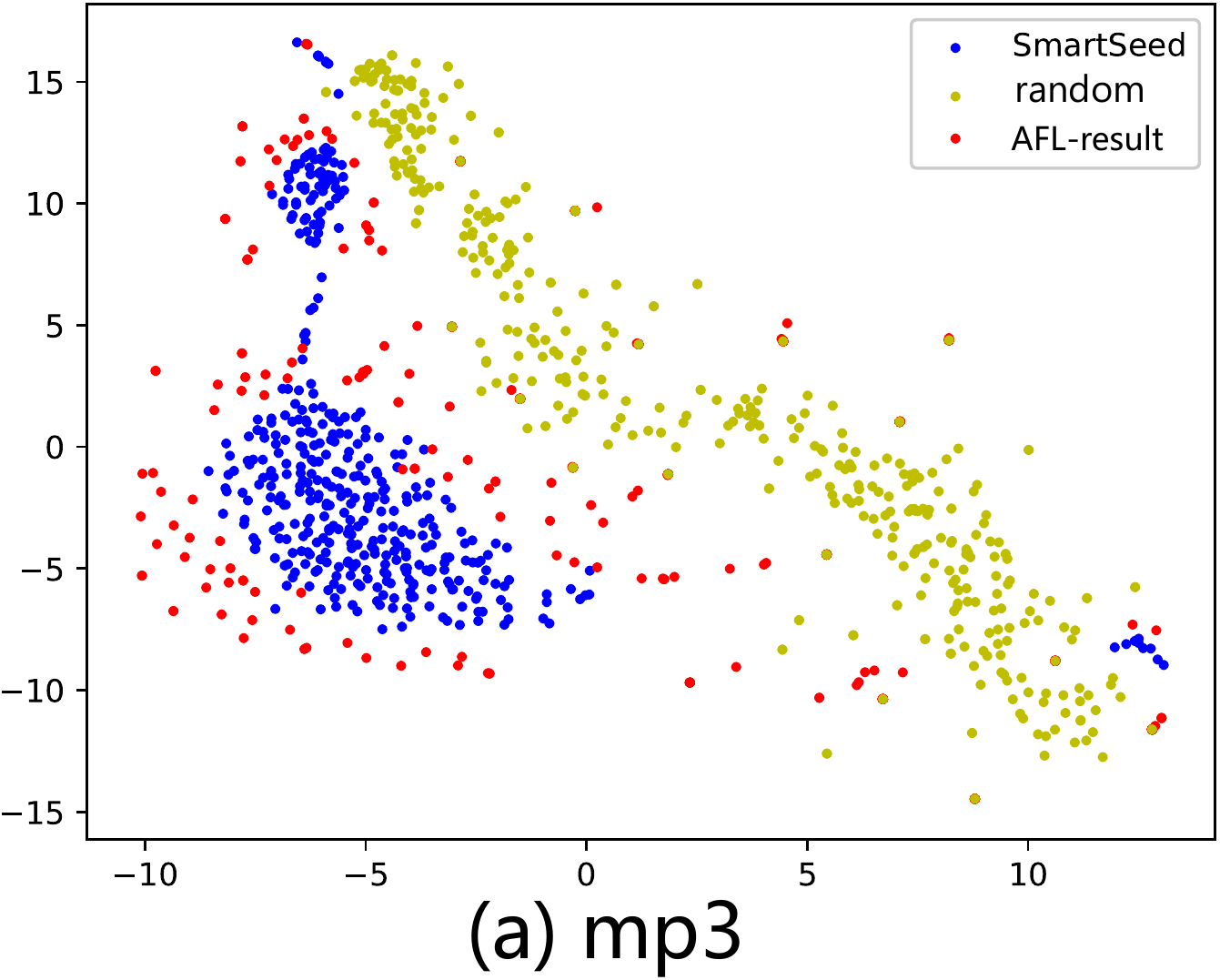}
\end{minipage}
\hfill
\begin{minipage}[t]{0.33\linewidth}
\centering
\includegraphics[height=3cm,width=0.8\textwidth]{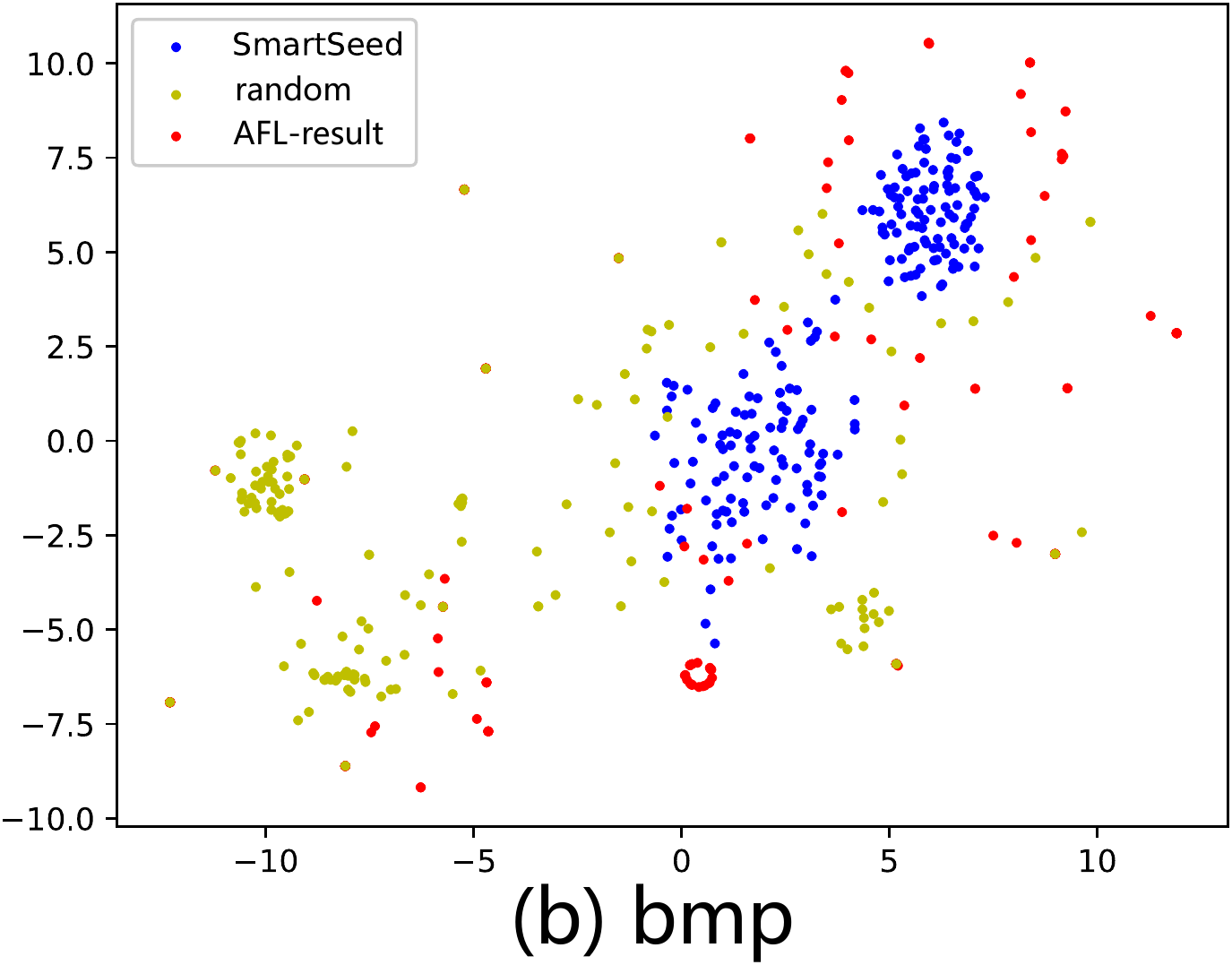}
\end{minipage}%
\hfill
\begin{minipage}[t]{0.33\linewidth}
\flushleft
\includegraphics[height=3cm,width=0.8\textwidth]{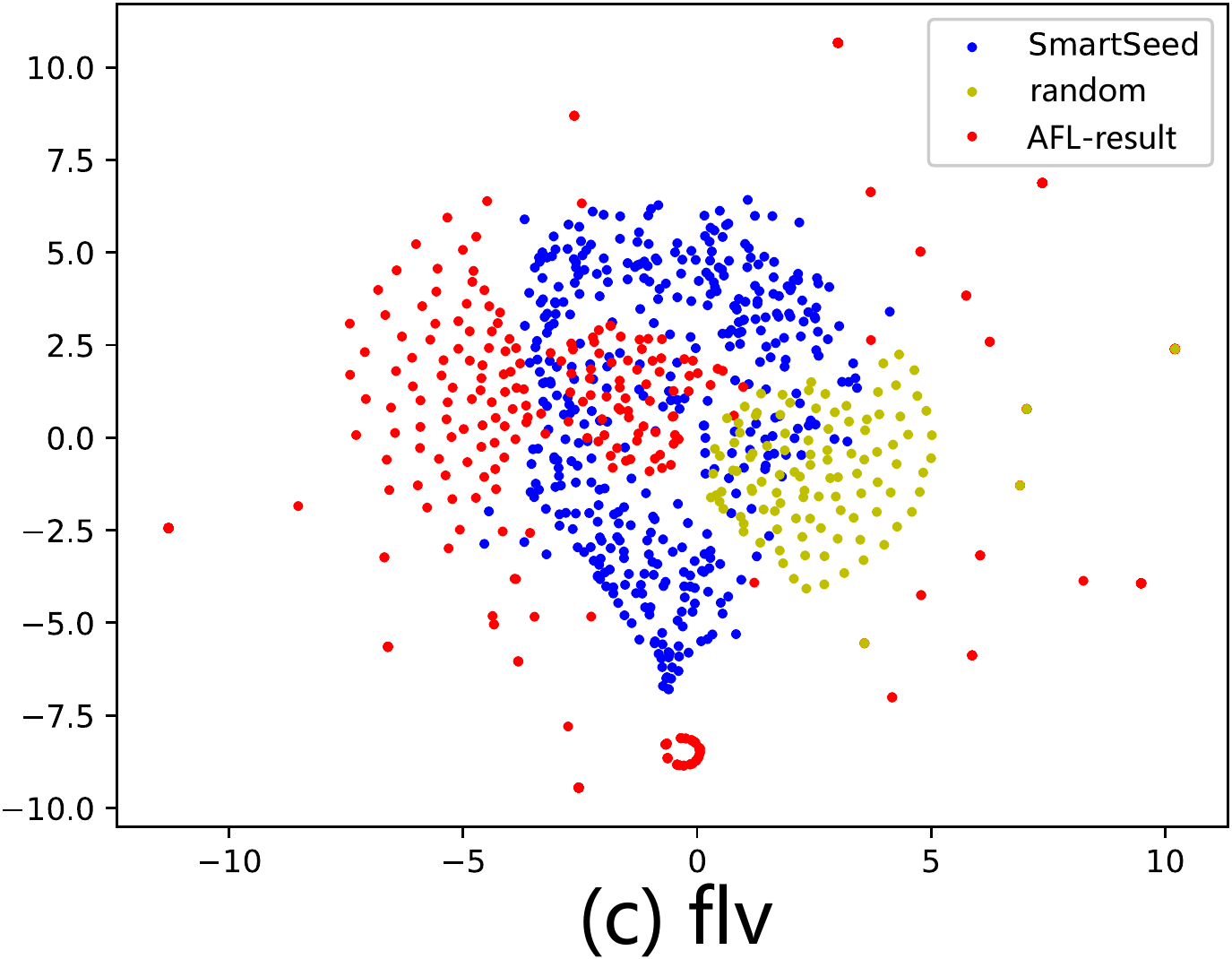}
\end{minipage}
\hfill
\caption{Similarity distribution of the seeds of \texttt{SmartSeed},   \texttt{random}  and  \texttt{AFL-result}. }\label{seeddistribution}
\end{center} 
\end{figure*}

\begin{figure*}  
\begin{center} 
\begin{minipage}[t]{0.329\linewidth}
\flushright
\includegraphics[height=3cm,width=0.8\textwidth]{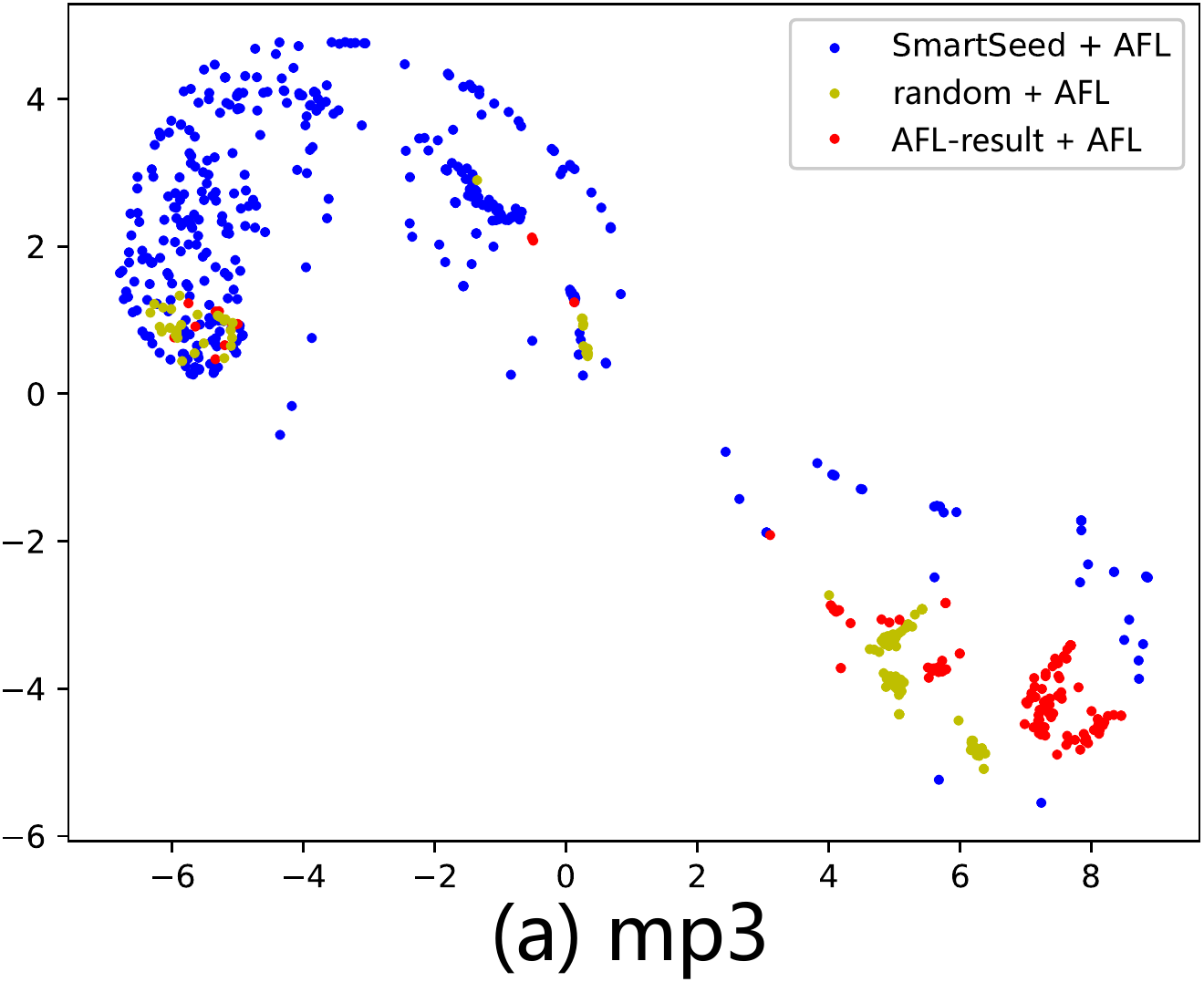}
\end{minipage}
\hfill
\begin{minipage}[t]{0.329\linewidth}
\centering
\includegraphics[height=3cm,width=0.8\textwidth]{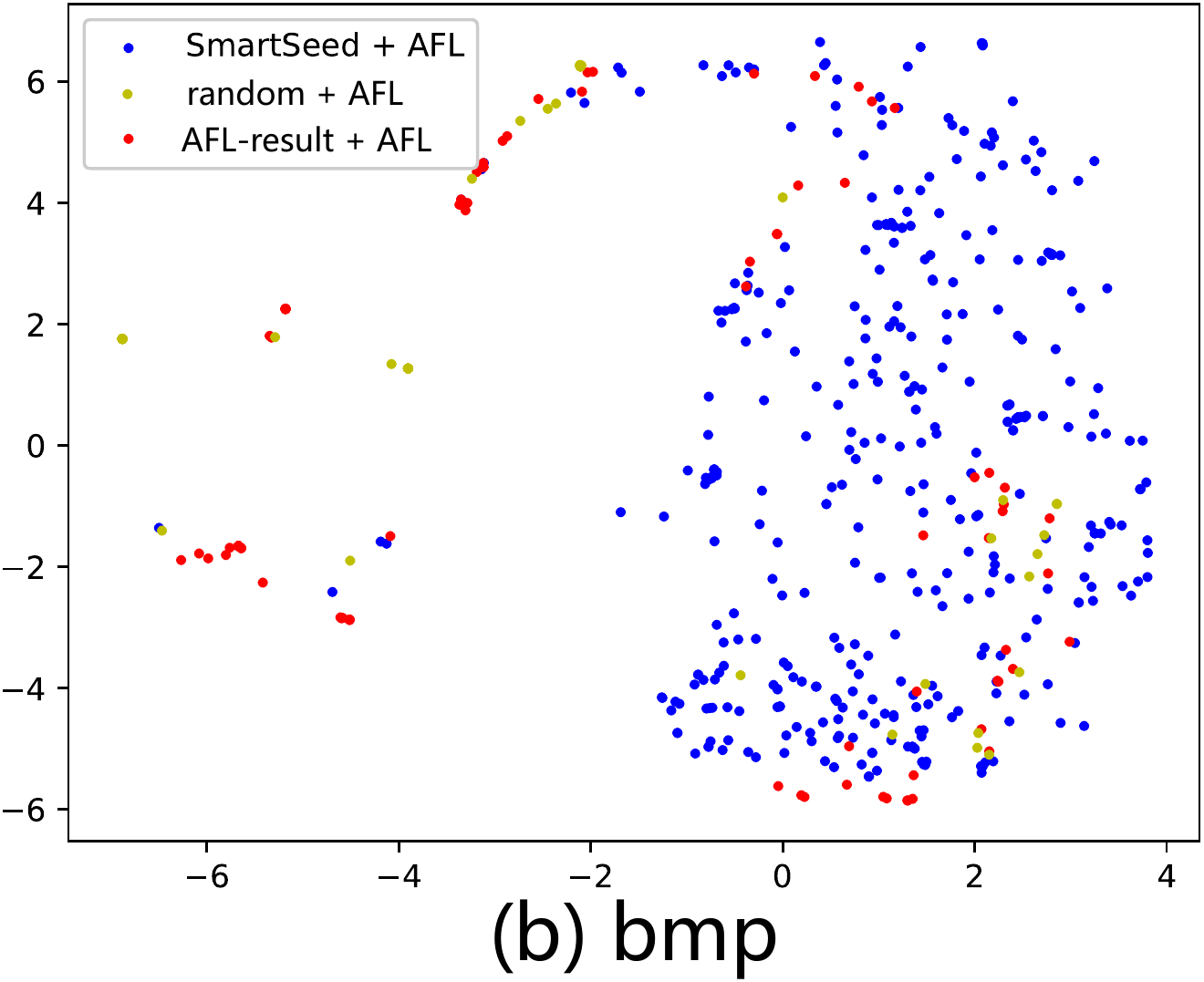}
\end{minipage}
\hfill
\begin{minipage}[t]{0.329\linewidth}
\flushleft
\includegraphics[height=3cm,width=0.8\textwidth]{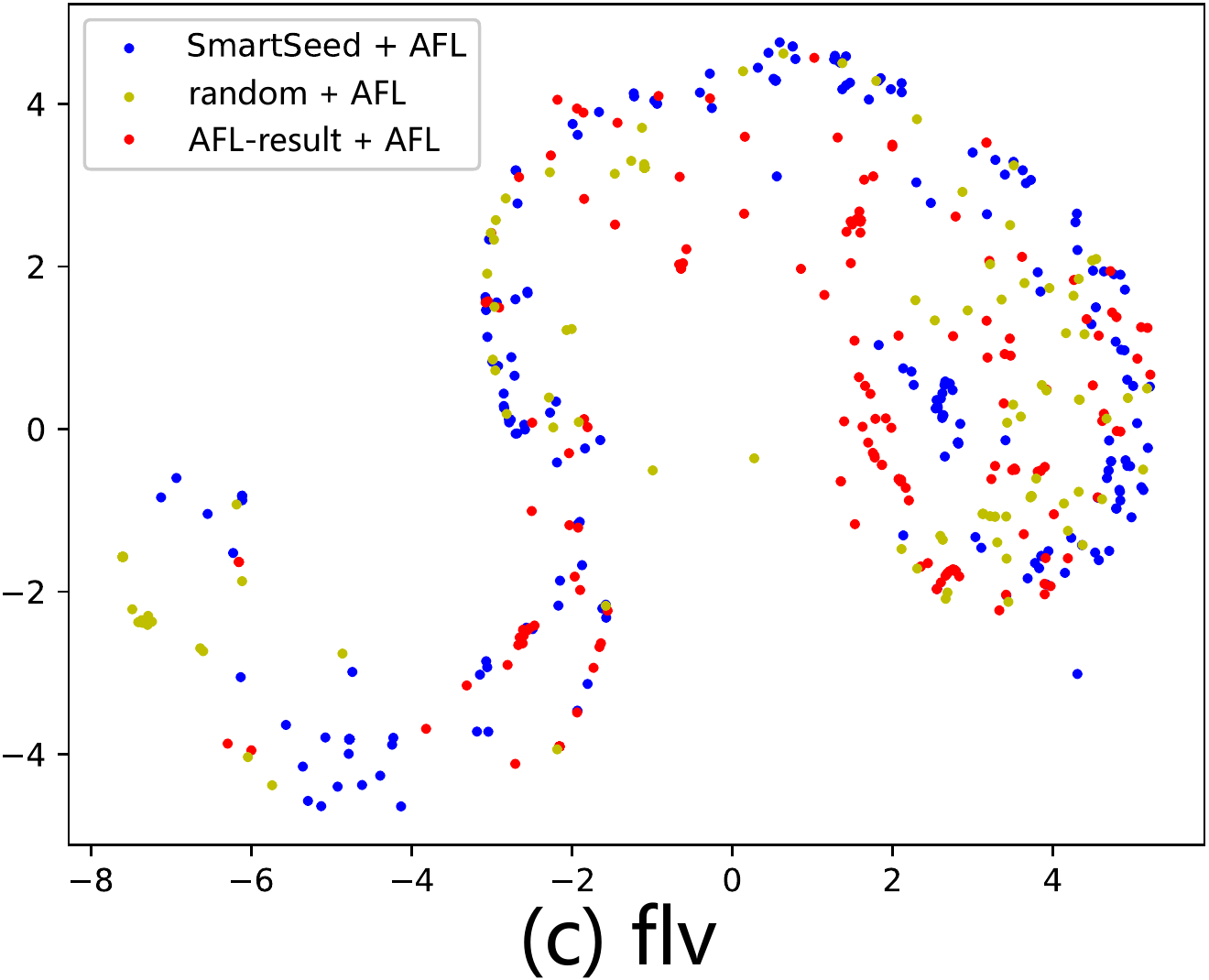}
\end{minipage}
\hfill
\caption{Similarity distribution of  valuable mutated input files of \texttt{SmartSeed}, \texttt{random}  and \texttt{AFL-result}   that triggered   crashes. }\label{crashdistribution}
\end{center} 
\end{figure*}

%To further understanding the effectiveness of \texttt{SmartSeed}, we make more in-depth analysis in this section. 

\subsection{Distribution}

We would like to examine the distribution of the seeds generated by different strategies, and the   distribution of the valuable files mutated from those seeds that trigger   crashes.

To facilitate our analysis, we employ t-SNE \cite{item35}, which is one of the best dimensionality reduction  algorithms that can aggregate similar data together, to visualize the distribution.  
Specifically, we use \texttt{SmartSeed}, \texttt{random} and \texttt{AFL-result}  to generate  300 seeds,  respectively, and then  visualize the distribution of these seeds  in Fig. \ref{seeddistribution}. Here, note that (1) the seeds of \texttt{AFL-result}  are selected from the stored files of AFL that can trigger crashes  or new paths, which are actually the mutated files of the \texttt{random}'s seeds  using AFL as we described  before   (the default seed selection strategy of AFL is \texttt{random} in our implementation); and (2) we do  not consider \texttt{peachset},  \texttt{hotset}, and \texttt{AFL-cmin}. This is mainly because  similar to  \texttt{random},   the seeds generated by these three schemes highly depend on the initial seed source dataset. It turns  out that the seeds obtained by them exhibit similar distribution as \texttt{random}. Thus, we do not plot their  distribution in Fig. \ref{seeddistribution} in order to make it more readable.     
From Fig. \ref{seeddistribution},  we have the following observations.    

(1)  
Although \texttt{AFL-result} is mutated from \texttt{random}, its distribution    is far away from  \texttt{random}. In other words, it takes time for \texttt{random} to discover valuable   input files that trigger   crashes or new paths. Compared with the distribution of \texttt{random}, the distribution of  \texttt{SmartSeed} is closer to \texttt{AFL-result}. This implies that  \texttt{SmartSeed} can discover   crashes  and paths faster than \texttt{random}.

(2) 
From Fig. \ref{seeddistribution}, 
we learn that   the distribution of \texttt{AFL-result} is more decentralized.  Compared with   \texttt{AFL-result}, the distribution of \texttt{SmartSeed} is more  intensive. \texttt{SmartSeed}  is also closer to the main  part of \texttt{AFL-result}.  
These facts indicate that:   
when using \texttt{AFL-result} to fuzz an application,  an   input file may spend a longer time to be   mutated into  another valuable input file because of the discrete distribution.  By   contrast, \texttt{SmartSeed} may be   mutated fast  into the main part of valuable input files that can trigger   crashes and paths of the objective  application.   
Thus, \texttt{SmartSeed} is more effective compared with \texttt{AFL-result}.  
%, while the  decentralized input files of \textbf{AFL-result} still try to be mutated into other valuable   files. 

Then, we leverage t-SNE to analyze the valuable mutated files of  \texttt{SmartSeed + AFL}, \texttt{random + AFL} and \texttt{AFL-result + AFL}, i.e., the  files are mutated from the seeds of \texttt{SmartSeed},   \texttt{random} and \texttt{AFL-result} that    trigger unique crashes of objective applications.  
Since \texttt{SmartSeed} discovers more crashes than \texttt{random} and \texttt{AFL-result}, its points are more than those two. Note that the distribution of valuable files discovered by \texttt{SmartSeed} will only be more sparse if we use the same number of files to plot.  
%The results are shown in Fig.7, from which we have the following observations. 
%Then, we use t-SNE to analyze the distribution of input files discovered by \texttt{SmartSeed + AFL}, \texttt{AFL-result + AFL} and \texttt{random + AFL} that found the crashes of the objective softwares. 
The  results  are  shown   in  Fig. \ref{crashdistribution},  from which we  have the following observations. 

(1)  
The distribution of  valuable  files mutated from \texttt{AFL-result} and \texttt{random} are similar. 
Also, it seems difficult for the seeds of \texttt{AFL-result} and \texttt{random} to be mutated into the distant  points that can trigger   crashes, which then limits  their fuzzing performance.  

(2)  On the contrary,  the points of the  valuable mutated files of \texttt{SmartSeed}   spread in a  larger area.  This demonstrates that the seed files generated by \texttt{SmartSeed} are easier to be mutated into the discrete  valuable input files that can trigger   crashes.

(3) 
We can learn from Fig. \ref{seeddistribution} and   Fig. \ref{crashdistribution}  that although the distributions  of \texttt{AFL-result} and \texttt{random} are more discrete than \texttt{SmartSeed}, the  valuable input files  discovered by \texttt{SmartSeed}, which trigger   unique crashes of   objective applications, cover  a larger area. In  other words, the relatively-concentrate distribution of the seeds generated by \texttt{SmartSeed} does not limit  them to be mutated into discrete input files and trigger more unique crashes.

In summary,  the distribution of \texttt{SmartSeed} is dense and closer to the main part of \texttt{AFL-result}.  Meanwhile, the seeds of \texttt{SmartSeed} seem   easier to be mutated into   valuable input files that can  trigger crashes.  %As for Fig. \ref{crashdistribution},  we learn that \texttt{SmartSeed} are easier to be mutated into discrete valuable files than \texttt{AFL-result} and \texttt{random}.  
These  observations may explain the better performance of \texttt{SmartSeed} from another angle.

\section{Limitations and Future Work.}   
As the focus of this paper, \texttt{SmartSeed} in its current form is designed for genetic algorithm based  fuzzing. Therefore, it is not suitable to use \texttt{SmartSeed} for generating highly-structured input files. We  take this issue as the future work of this paper and keep extending our system. 
From the performance perspective, \texttt{SmartSeed} could be improved from many aspects. For instance, a  better generative model could be designed. Also, it is interesting to study the best working scope of different  generative models and how to further improve the model training process.  
From the evaluation perspective, our primary goal in this paper is to demonstrate the performance and usability  of \texttt{SmartSeed}. Certainly, more evaluations can be conducted to comprehensively evaluate  \texttt{SmartSeed}, e.g., considering more applications and more formats.

\section{Related Work}\label{section6}
In this section, we briefly  introduce the related work. 

\textbf{Mutation-based Fuzzing.}   
As a representative of mutation-based fuzzing,   AFL \cite{item11} employs a novel type of compile-time  instrumentation and   genetic algorithms to automatically discover the valuable input files that trigger   new  paths or   unique crashes. Because of its high-efficiency and ease of use, AFL is one of the most popular fuzzing  tools. 
Based on AFL,  B\"{o}hme et al.  presented a fuzzing tool named \emph{AFLFast} that can detect   more  paths   by prioritizing low-frequency paths \cite{item2}.   
%B\"{o}hme, Pham and Roychoudhury observed that most input files test the few high-frequency paths \cite{item2}. Thus, based on AFL, they presented the fuzzing tool named \emph{AFLFast} that detects distinctly more paths with the same number of input files by guiding towards low-frequency paths. %The result showed that \emph{AFLFast} discovers much more unique crashes and paths than AFL. 
In \cite{item40}, B\"{o}hme  et al.    combined a simulated annealing-based power schedule scheme with AFL and presented \emph{AFLGo}.  
% Kersten et al.  presented a tool named \emph{Kelinci}   to  interface AFL with instrumented Java based  softwares \cite{item42}.   
Xu  et al.  implemented three new operating primitives  specialized for improving the fuzzing  performance of large-scale tasks on multi-core machines  \cite{item43}.   
%we design and implement three new operating primitives  specialized for fuzzing that solve these performance bottlenecks  and achieve scalable performance on multi-core machines.
In \cite{item36}, Gan et al. presented  a  solution to mitigate  path collisions, which can be combined with AFL and AFLFast to construct CollAFL and CollAFL-fast, respectively.

Many  other  works focus on combining fuzzing with other bug detection technologies such as   taint tracing, symbolic execution and program analysis. 
%Ganesh et al. presented an automated white box fuzzing tool named \emph{BuzzFuzz}, which uses dynamic taint tracing to locate the ranges of valuable bytes in the input files \cite{item21}. 
Wang et al. combined   fuzzing   with dynamic taint analysis and symbolic execution techniques and presented a fuzzing tool named \emph{TaintScope} \cite{item22}. Haller et al. presented a   fuzzing tool named \emph{Dowser}, which takes   taint tracking, program analysis and symbolic execution into consideration \cite{item7}.   
Sang et al. considered to employ white-box symbolic analysis in their fuzzing tool design \cite{item4}. 
Stephens et al. also   considered  to    involve selective symbolic execution into their   fuzzing tool named \emph{Driller}   \cite{item5}. 
Pham et al. combined input model-based approaches  with symbolic execution and  presented Model-based Whitebox Fuzzing (MoWF) \cite{item23}.

To improve the coverage, 
Rawat et al. presented an application-aware evolutionary fuzzing tool named \emph{VUzzer} \cite{item3}. \emph{VUzzer} uses static and dynamic analysis to analyze the priority of  paths. 
In \cite{item45},   Peng  et al. presented  \emph{T-Fuzz},  which uses a dynamic tracing based technique to  detect and   remove the checks in   objective applications  to improve the code coverage.  
By using scalable byte-level taint tracking, context-sensitive branch count,       gradient descent  based  search,  shape and type inference and input length exploration to solve path constraints,    Angora  presented by Chen et al.   can increase  the branch coverage of objective applications \cite{item46}.

Note that \texttt{SmartSeed} is designed as a generic system to generate valuable  
seed files for and to be easily compatible with   mutation-based fuzzing. With the seeds generated by   \texttt{SmartSeed}, we expect to   improve the performance of mutation-based fuzzing.

\textbf{Generation-based Fuzzing.}   
Generation-based fuzzing tools are designed to  generate   input files with   specific input formats.  %themselves. Thus, it seems hard to combine our system with the generation-based fuzzing tools. 
Following this track, 
Godefroid  et al. used the grammar-based specification of   valid highly-structured input files to improve the  performance of fuzzing     \cite{item19}.  
Holler et al. presented \emph{LangFuzz} to fuzz the applications with   highly-structured inputs   such as   JavaScript interpreters \cite{item24}. 
In order to fuzz   compilers and interpreters, 
Dewey et al. proposed to use Constraint Logic Programming (CLP) for the program generation  \cite{item25}. %With CLP, Users can manually write declarative predicates to specify interesting programs 
Recently, 
Wang et al. presented  a novel data-driven seed generation approach named \emph{Skyfire}  \cite{item6},  which   uses   Probabilistic Context-Sensitive Grammar (PCSG) to learn the syntax features and semantic rules from the training set. %Then, it generates input files for the fuzzing tools such as AFL. 
%2017
In  \cite{item31}, Godefroid et.al. presented a RNN-based  machine-learning technique  to generate highly-structured format files such as PDF. The method not only can pass the format checks with a high probability, but also can improve the code coverage. 
%2018

The primary difference  between \texttt{SmartSeed} and generation-based fuzzing approaches is that our method is used to generate   binary seed files such as image, music and video, while they focus on improving the fuzzing efficiency of  applications with   highly-structured input formats. 

\textbf{Other Fuzzing Strategies.}   
As for kernel vulnerabilities,  
 Corina et al. presented an interface-aware fuzzing tool named \emph{DIFUZE} to automatically generate  inputs for kernel drivers  \cite{item41}.  
 Han et al.   proposed a novel method called \emph{model-based API fuzzing} and presented  \emph{IMF} for testing commodity OS kernels  \cite{item44}.   
You et al.   presented a novel technique named \emph{SemFuzz}  \cite{item47}  that can learn from  vulnerability-related  texts such as CVE reports  and   automatically generate Proof-of-Concept (PoC) exploits.  
Petsios et al. focused on algorithmic  complexity vulnerabilities and proposed \emph{SlowFuzz}  to   generate inputs to  trigger the maximal   resource  utilization  behavior  of applications and algorithms   \cite{item48}.    

\textbf{Seed Selection.}   
To figure out how to select a better initial seed set, 
Allen and Foote presented an algorithm to consider the parameter selection and automated selection of seed files \cite{item30}.   %The algorithm was implemented in an open-source Basic Fuzzing  Framework (BFF). 
Woo et al. developed a framework to evaluate 26 randomized online scheduling algorithms to see which    can schedule   better seeds to fuzz a program \cite{item29}.  
%One of their new scheduling algorithms performed the best during their experiments. 
  Rebert et al. evaluated six seed selection strategies of fuzzing  and presented several interesting  conclusions about   seed sets \cite{item1}.   
They  also showed the necessity to select a good seed set to   improve the efficiency of   fuzzing.   
Recently, Nichols et.al.      showed that using the generated   files of GAN to reinitialize AFL can potentially find more unique paths of \texttt{ethkey}   \cite{item32}.  
However, they neither  described the model in detail   nor provided any prototype. Thus, we cannot reproduce their   model and compare it with \texttt{SmartSeed}.  
%It seems our system has better augmentability, since users  can adjustment the file size of generated seeds. We tested 12 softwares and proved that our generated seed set can discover more unique crashes and paths in the most cases. 

Different from existing research, we focus on designing a generic seed generation system leveraging start-of-the-art machine learning techniques.  We also demonstrate its effectiveness,  robustness  and   compatibility  through  extensive experiments.

\textbf{GAN Models.}   
In 2014, Goodfellow et al. presented a new unsupervised  learning framework named  Generative Adversarial Networks (GAN)    \cite{item8}.   
%In order to obtain the generative model that can generate verisimilar data, two models alternately train  each other and improve each other. 
To improve  GAN, Radford et al. tried multiple combinations of   machine learning models to construct better  generative and discriminative models for  GAN,  and presented the Deep Convolutional GAN (DCGAN) model   \cite{item9}.  
Later, Zhai et al. combined GAN with an Energy Based Model (EBM) and  proposed VGAN that works by minimizing a variational lower bound of the negative log likelihood of   EBM \cite{item38}. 
Chen et al. combined GAN with mutual information and    presented InfoGAN \cite{item39}, which can unsupervisedly learn  interpretable  representations. 
Recently, Arjovsky et al. presented the Wasserstein GAN (WGAN) model by using the approximation of the Wasserstein distance as the  loss function \cite{item10}. 

%Unlike other GAN models, WGAN   improves the stability of learning a lot. It is also much easier to train a WGAN model. In addition, WGAN   can solve the problems of GAN like mode collapse in most application scenarios. 

Since GAN and   its variants can use unsupervised learning methods to generated   more realistic   data,  they have been applied in many applications such as high-quality image  generation \cite{item52, item53, item54} and image translation \cite{item49, item50, item51}.  
In this paper, we extend the application of GAN to improve the performance of fuzzing.

% if have a single appendix:
%\appendix[Proof of the Zonklar Equations]
% or
%\appendix  % for no appendix heading
% do not use \section anymore after \appendix, only \section*
% is possibly needed

% use appendices with more than one appendix
% then use \section to start each appendix
% you must declare a \section before using any
% \subsection or using \label (\appendices by itself
% starts a section numbered zero.)
%

%\appendices

% trigger a \newpage just before the given reference
% number - used to balance the columns on the last page
% adjust value as needed - may need to be readjusted if
% the document is modified later
%\IEEEtriggeratref{8}
% The "triggered" command can be changed if desired:
%\IEEEtriggercmd{\enlargethispage{-5in}}

% references section

% can use a bibliography generated by BibTeX as a .bbl file
% BibTeX documentation can be easily obtained at:
% http://mirror.ctan.org/biblio/bibtex/contrib/doc/
% The IEEEtran BibTeX style support page is at:
% http://www.michaelshell.org/tex/ieeetran/bibtex/
%\bibliographystyle{IEEEtran}
% argument is your BibTeX string definitions and bibliography database(s)
%\bibliography{IEEEabrv,../bib/paper}
%
% <OR> manually copy in the resultant .bbl file
% set second argument of \begin to the number of references
% (used to reserve space for the reference number labels box)

\end{document}